\documentclass[aps,prd]{revtex4}
\usepackage{graphicx}
\usepackage{amsmath}
\usepackage[all]{xy}
\usepackage{amssymb}
\newcommand{\be}{\begin{equation}}
\newcommand{\ee}{\end{equation}}
\newcommand{\ben}{\begin{eqnarray}}
\newcommand{\een}{\end{eqnarray}}
\newcommand{\bes}{\begin{subequations}}
\newcommand{\ees}{\end{subequations}}
\newcommand{\bb}{\bibitem}
\newcommand{\ov}{\overline}
\newcommand{\wt}{\widetilde}

\begin{document}
\title{Defect Structures in Field Theory}
\author{Dionisio Bazeia}
 \address{Departamento de F\'\i sica, Universidade Federal da Para\'\i ba\\
  58051-970 Jo\~ao Pessoa, Para\'\i ba, Brazil}

\begin{abstract}
{This work contains a set of lectures on defect structures, mainly in models described by scalar fields in diverse dimensions.}
\end{abstract}
\maketitle
%%%%%%%%%%%%%%%%%%%%%%%%%%%%%%%%%%%%%%%
%%%%%%%%%%%%%%%%%%%%%%%%%%%%%%%%%%%%%%%
%%%%%%%%%%%%%%%%%%%%%%%%%%%%%%%%%%%%%%%
\bigskip
\section{Introduction}
\bigskip

This work describes a set of lectures on {\it Defect Structures in Field Theory.} Although we focus mainly on issues concerning scalar fields, we also deal a little with other topics. We work mainly in $(1,1)$ dimensions, in flat space-time with metric such that $diag(\eta_{\mu\nu})=(1,-1),$ and we set the Planck constant $\hbar$ and the speed of light $c$ to unit, and we use dimensionless fields and coordinates. 

The interested reader can find motivations and general investigations in the books \cite{B1,B2,B3}, where the authors explore several aspects of defect structures such as kinks, vortices and monopoles. There is another book, Ref.~\cite{B4}, in which the reader may read important original articles in the subject until 1984.

%%%%%%%%%%%%%%%%%%%%%%%%%%%%%%%%%%%%%%%%%%%%%%%%%%%%%%%%%%%
%%%%%%%%%%%%%%%%%%%%%%%%%%%%%%%%%%%%%%%%%%%%%%%%%%%%%%%%%%%
%%%%%%%%%%%%%%%%%%%%%%%%%%%%%%%%%%%%%%%%%%%%%%%%%%%%%%%%%%%
\section{One-field models}
\bigskip

We start with
\be
{\mathcal L}=\frac12\partial_\mu\phi\partial^\mu\phi-V(\phi)
\ee
where $V(\phi)$ is the potential, which specifies the model under consideration.

We are working in $(1,1)$ dimensions, and so the equation of motion which follows from the above model is
\be\label{em}
\frac{\partial^2\phi}{\partial t^2}-
\frac{\partial^2\phi}{\partial x^2}+\frac{dV}{d\phi}=0
\ee
If the field is constant we get $dV/d\phi=0.$ However, for static field configuration we have
\be\label{ems}
\frac{d^2\phi}{dx^2}=\frac{dV}{d\phi}
\ee
We are interested in solutions with finite energy. For the trivial constant solutions we have
\be
{E_0}=\int_{-\infty}^{\infty} dx\,V(\bar\phi)
\ee
and for static solutions 
\be
E=\int_{-\infty}^{\infty}dx\,\biggl[\frac12\left(\frac{d\phi}{dx}\right)^2+
V(\phi)\biggr]
\ee
In this last expression, the two terms identify the gradient and potential portions of the total energy, respectively.

The trivial constant solution is represented by $\bar\phi.$ It obeys $V^\prime(\bar\phi)=0$ and we should also impose that $V(\bar\phi)=0,$ to make $\bar\phi$ to describe a null energy solution. On the other hand, static solutions must obey the boundary conditions
\be\label{cc1}
\lim_{x\to-\infty}\frac{d\phi}{dx}\to0
\ee
and
\be\label{cc2}
\lim_{x\to-\infty}\phi(x)\to \bar\phi
\ee
These conditions are necessary to make both the gradient and potential energies finite, independently.
This behavior can be verified with the equation of motion: we rewrite Eq.~(\ref{ems}) to get
\be
\frac{d\phi}{dx}=\pm\sqrt{2V}
\ee
which also shows that the potential for a static solution cannot be negative.

The behavior of the solution in the limit $x\to\infty$ may be equal to the former case, that is, $\phi(x\to\infty)\to \bar\phi;$
or it may be different, giving $\phi(x\to\infty)\to{\bar\varphi},$ for $\bar\varphi$ being another trivial solution
that obeys $V^\prime(\bar\varphi)=0,$ with $V(\bar\varphi)=0.$ These two possibilities can be described by the topological current
\be
j^{\mu}_T=\frac12\varepsilon^{\mu\nu}\partial_\nu\phi
\ee
which is conserved, that is, which obeys $\partial_\mu j^\mu_T=0.$ For static solutions the topological charge is given by
\be
Q_T=\frac12\phi(x\to\infty)-\frac12\phi(x\to-\infty)
\ee 
This charge sees the asymptotic bahavior of the static solutions, and it specifies two distinct cases:
topological solutions, for which $\phi(x\to-\infty) \neq \phi(x\to\infty),$ and non-topological solutions,
for which $\phi(x\to-\infty)=\phi(x\to\infty).$ Thus, in general we can have topological or kink-like solutions, and non-topological or lump-like solutions. We can also have ribbons, which are kinks embedded in $(2,1)$ dimensions, and walls, which are kinks embedded
in $(3,1)$ dimensions.

We notice that there are several distinct ways to define the topological current, and this can be used to define the topological charge appropriately.

In the case of a single real scalar field we can define the standard model, which engenders spontaneous symmetry breaking, with the potential
\be\label{pphi4}
V(\phi)=\frac12\,(1-\phi^2)^2
\ee
This is the $\phi^4$ model, and it is plotted in Fig.~1.

The equation of motion is
\be
\frac{\partial^2\phi}{\partial t^2}-\frac{\partial^2\phi}{\partial x^2}+
2\phi(\phi^2-1)=0
\ee
and for static solutions we get
\be
\frac{d^2\phi}{dx^2}=2\phi(\phi^2-1)
\ee
We are searching for extended but localized solutions, with finite energy. The equation above has two trivial solutions,
given by $\bar\phi_{\pm}=\pm1,$ which have zero energy and identify the classical vacua on the model. There are static solutions,
given by 
\be
\phi_{\pm}(x)=\pm\,\tanh(x)
\ee
They identify the kink ($+$) and anti-kink ($-$), with center at the origin, $x=0$. In fact, the center of a static solution
is arbitrary, because the model has translational invariance.

%%%%%%%%%%%%%%%%%%%%%%%%%%%%%%%%%%%%%%%%%%%%%%%%%%%%%%%%%%%%%%%%%
\begin{figure}
\begin{center}
\includegraphics[height=6cm,width=10cm]{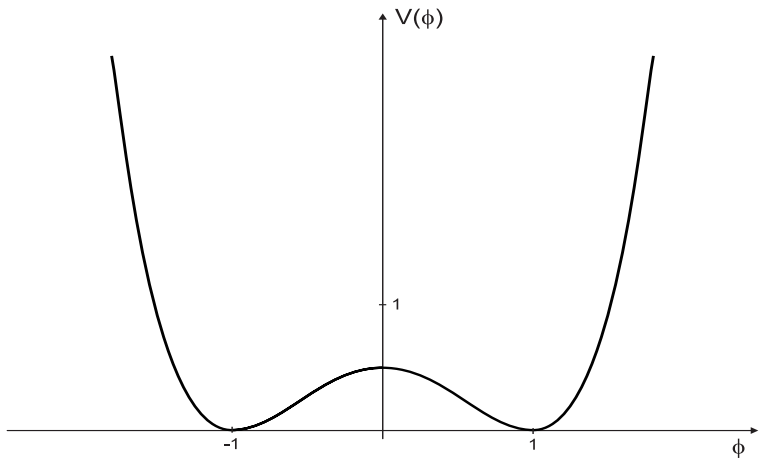}
\end{center}
\caption{Potential for the $\phi^4$ model}
\end{figure}
%%%%%%%%%%%%%%%%%%%%%%%%%%%%%%%%%%%%%%%%%%%%%%%%%%%%%%%%%%%%%%%%% 

The kink above has amplitude and width equal to unit. However, if we write the potential in the form
\be
V(\phi)=\frac12\lambda^2(a^2-\phi^2)^2
\ee
the kink would be $\phi(x)=a\tanh(\lambda a x),$ and it would have amplitude $a$ and width inversely proportional to $\lambda a.$  

The energy for static solutions has the form
\be
E=\int_{-\infty}^{\infty} dx\; \varepsilon(x)
\ee
where $\varepsilon(x)$ is the energy density, which can be written as 
\be
\varepsilon(x)=\frac12\left(\frac{d\phi}{dx}\right)^2+V(\phi)
\ee
We see that for the trivial solution ${\bar\phi},$ which satisfies $V(\bar\phi)=0,$ the energy is zero. For the kink or anti-kink
the energy is $E=4/3$, and the energy density is
\be\label{ekink}
\varepsilon(x)={\rm sech}^4(x)
\ee
We notice that the gradient and potential portions of energy are given by
\be
E_g=\frac12\int_{-\infty}^{\infty}dx \left(\frac{d\phi}{dx}\right)^2
\ee
and 
\be
E_p=\int_{-\infty}^{\infty}dx\, V(\phi)
\ee
They vanish for the trivial solutions $\bar\phi_\pm=\pm1.$ For the kink we get
\be
\varepsilon_p(x)=\varepsilon_g(x)=\frac12 {\rm sech}^4(x)
\ee
which shows that $E_g=E_p.$ 

We can consider another model, described by the potential
\be\label{pphi4l}
V(\phi)=\frac12 \phi^2-\frac12\phi^4
\ee
which is plotted in Fig.~2. We see that this new potential is similar to the former one, but inverted; thus we name it the inverted $\phi^4$ model.

%%%%%%%%%%%%%%%%%%%%%%%%
\begin{figure}
\begin{center}
\includegraphics[height=6cm,width=10cm]{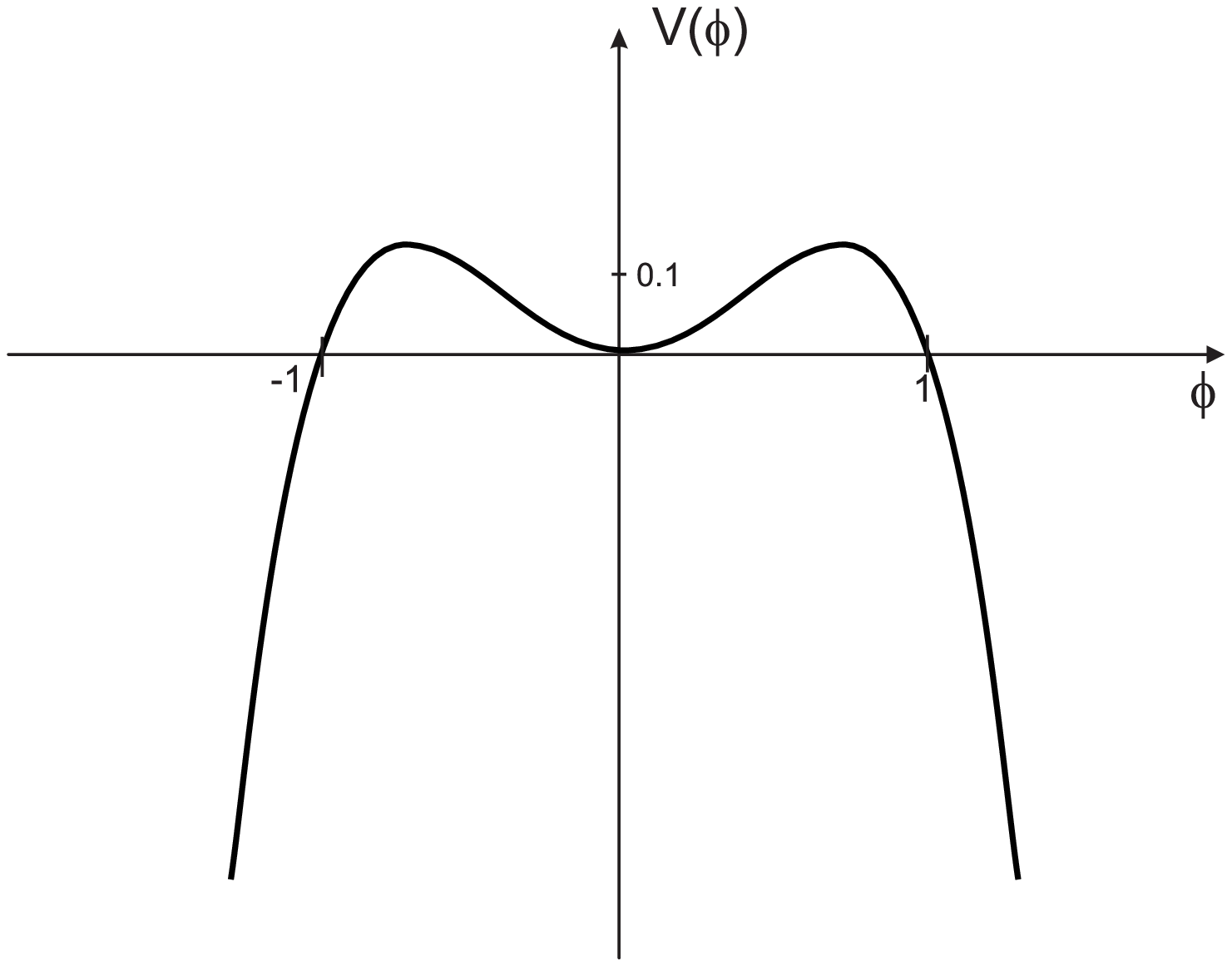}
\end{center}
\caption{Potential for the inverted $\phi^4$ model.}
\end{figure}
%%%%%%%%%%%%%%%%%%%%%%%%%%%%%%%%

In this case, the equation of motion for static solutions is
\be
\frac{d^2\phi}{dx^2}=\phi-2\phi^3
\ee
There is a single trivial solution, $\bar\phi=0,$ for which $V(\bar\phi)=0.$ And there are static solutions, given by
\be
\phi_\pm(x)=\pm\,{\rm sech}(x)
\ee
They are lumps, and they have energy density in the form
\be\label{elump}
\varepsilon(x)={\rm sech}^2(x)\tanh^2(x)
\ee
In Fig.~3 we plot kink and lump, to compare their profile.

%%%%%%%%%%%%%%%%%%%%%%%%%%%%%%%%%%%%%%%%%
\begin{figure}[ht]
\begin{center}
\includegraphics[height=6cm,width=10cm]{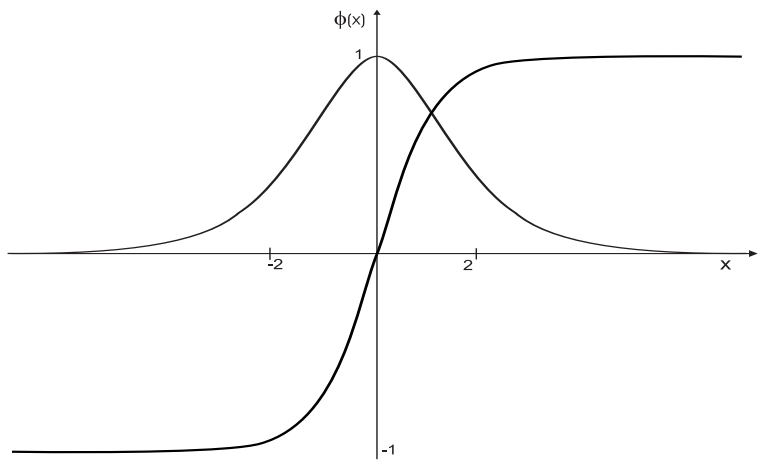}
\end{center}
\caption{Plots of the solutions for kinks and lumps.}
\end{figure}
%%%%%%%%%%%%%%%%%%%%%%%%%%%%%%%%%%%%%%%%% 

%%%%%%%%%%%%%%%%%%%%%%%%%%%%%%%%%%%%%%%%%
\begin{figure}[ht]
\begin{center}
\includegraphics[height=6cm,width=10cm]{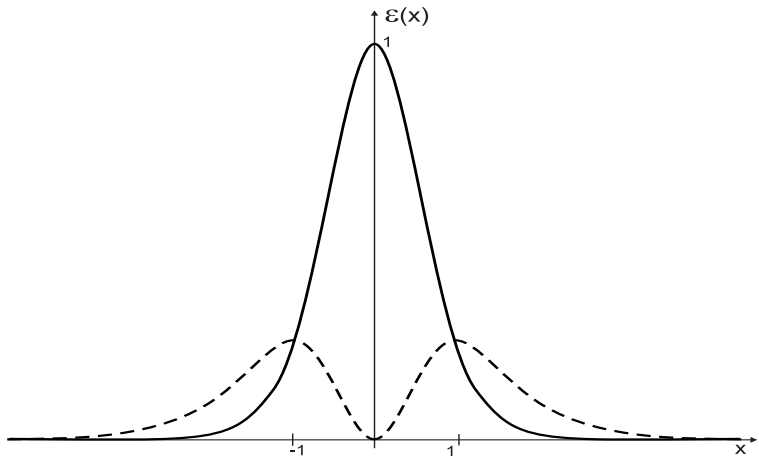}
\end{center}
\caption{Energy density for kinks and lumps, depicted as solid and dashed lines, respectively.}
\end{figure}
%%%%%%%%%%%%%%%%%%%%%%%%%%%%%%%%%%%%%%%%% 

This last result shows that the energy density of the lump vanishes at the origin, the center of the lump, and is maximized at the points
$x=\pm{\rm arctanh}(\sqrt{2}/2)\approx\pm0.88.$ This suggests that lumps may have internal structure. In Fig.~4
we plot the energy density of both kink and lump, as given by Eq.~(\ref{ekink}) and (\ref{elump}). There we see that
the energy density of the kink gets concentrated at the origin, but this is not the case for lumps.

%%%%%%%%%%%%%%%%%%%%%%%%%%%%%%%%%%%%%%%%%%%%%%%
%%%%%%%%%%%%%%%%%%%%%%%%%%%%%%%%%%%%%%%%%%%%%%%
%%%%%%%%%%%%%%%%%%%%%%%%%%%%%%%%%%%%%%%%%%%%%%%
\subsection{BPS States}

There is an alternative way to search for defect solutions. It is due to Bogomol'nyi \cite{B} and Prasad and Sommerfield \cite{PS}.
The point is that for non-negative potentials we can write
\be\label{pw}
V(\phi)=\frac12 W_\phi^2
\ee
where $W=W(\phi)$ is a smooth function of the scalar field, and $W_\phi=dW/d\phi.$
In this case the equation of motion for static solutions is
\be
\frac{d^2\phi}{dx^2}=W_{\phi}\,W_{\phi\phi}
\ee
It is possible to see that solutions of the first-order equations
\be\label{eb}
\frac{d\phi}{dx}=\pm W_\phi
\ee
solve the second-order equation of motion. The solutions of the first-order equations are named BPS states,
and it is possible to show that they are stable, minimum energy static solutions. In the present case,
it is possible to show that the model only supports BPS states  \cite{bms1,bms2}.

To see how the BPS states appear, let us investigate the energy of the static solutions. We have 
\be
E=\frac12\int_{-\infty}^{\infty}dx\,
\biggl[\left(\frac{d\phi}{dx}\right)^2+W^2_\phi\biggr]
\ee 
We can write
\be
E=E_{BPS}+\frac12\int_{-\infty}^{\infty}dx\,
\left(\frac{d\phi}{dx}\mp W_\phi\right)^2
\ee
where
\be
E_{BPS}={||}\,{W[\phi(\infty)]-W[\phi(-\infty)]}\,{||}
\ee
is the bound in energy, which is attained for solutions that obey the first-order Eqs. (\ref{eb}).

The BPS states present interesting properties. First, they are solutions of first-order equations, and their energies are easily obtained, since they only depend on the $W(\phi),$ and on the asymptotic behavior of the field configuration. Second, the gradient and potential portions of their energies are equal, that is $E_g=E_p=(1/2)E_{BPS}.$ Another fact is that the BPS states appear very naturally
in the supersymmetric environment, and there they partially preserve the supersymmetry. In the presence of supersymmetry, $W$ is named superpotential; see Refs.~\cite{bag,mba,etb,nie,jaf,shi1} for more details on the subject. 

There are different motivations to investigate models which support BPS states. In particular,
in Ref.~\cite{ela} one investigates elastic deformations in liquid crystals, and there the decomposition in terms of BPS states
has led to wide class of solutions. Another interesting work was recently done in supersymmetric quantum mechanics \cite{mq}.
In this case, it was shown how to understand shape invariance in terms of the BPS structure of the system.

%%%%%%%%%%%%%%%%%%%%%%%%%%%%%%%%%%%%%%%%%%%%%%%%%%%%%%%
%%%%%%%%%%%%%%%%%%%%%%%%%%%%%%%%%%%%%%%%%%%%%%%%%%%%%%%
%%%%%%%%%%%%%%%%%%%%%%%%%%%%%%%%%%%%%%%%%%%%%%%%%%%%%%%
\subsection{Linear Stability}

Linear or classical stability of static solutions are investigated as follows: for $\phi=\phi(x)$ being static solution,
we can write
\be
\phi(x,t)=\phi(x)+\eta(x,t)
\ee
where $\eta(x,t)$ describes fluctuations around the solution $\phi(x).$ We suppose that the fluctuations are small,
and we only consider linear contributions in $\eta.$ We then use the above configuration and the equation of motion
(\ref{em}) to get, up to first-order in $\eta,$
\be
\frac{\partial^2\eta}{dt^2}-\frac{\partial^2\eta}{\partial x^2}+
\frac{d^2V}{d\phi^2}\Biggr|_{\phi(x)}\,\eta=0
\ee
Since the static solution only depend on $x,$ we can consider
\be\label{flu}
\eta(x,t)=\sum_n \eta_n(x)\cos(w_n t) 
\\\ee
to obtain the Schr\"odinger-like equation
\be\label{es1}
H\,\eta_n(x)=w^2_n\,\eta_n(x)
\ee
where
\be\label{es2}
H=-\frac{d^2}{dx^2}+\frac{d^2V}{d\phi^2}\Biggr|_{\phi(x)}
\ee

In the Schr\"odinger-like equation, we notice that the presence of negative eigenvalue transforms the $\cos$ factor in (\ref{flu}) into a $\cosh$, and this would violate the supposition that we are dealing with small fluctuations. Thus, we need that the associated quantum-mechanical problem have no negative eigenvalue, to make the solution stable.

An interesting fact is that the associated quantum-mechanical problem always contains a zero-mode, that is, an eigenfunction with zero eigenvalue. We prove this assertion by noting that if $\phi(x)$ is a static solution, we use the equation of motion and $\phi^\prime=d\phi/dx$ to obtain
\be
-\frac{d^2\phi^{\prime}}{dx^2}+\frac{d^2V}{d\phi^2}\,\phi^\prime=0
\ee
which is a Schr\"odinger-like equation with zero eigenvalue. Thus, we identify the zero-mode with the derivative
of the classical solution itself. The presence of the zero mode is a general fact, which appear in decorrence of the translational
invariance of the model under investigation \cite{jac}. For instance, for the $\phi^4$ model the kinks are
$\phi_{\pm}(x)=\pm\tanh(x).$ They are stable, and the zero-modes are $\phi_{\pm}^\prime(x)=\pm {\rm sech}^2(x),$ as we are going to show below. For the inverted $\phi^4$ model, the lumps are $\phi_{\pm}(x)=\pm{\rm sech}(x).$ They are unstable, with zero-modes
as $\phi_{\pm}^\prime(x)=\mp\tanh(x)\,{\rm sech(x)}.$ Their instabilities appear from the fact that such zero-modes cross zero
at the origin, so they have nodes and cannot represent lowest bound states \cite{landau}.

In the case of the non-negative potentials described in (\ref{pw}), there are BPS states, which obey first-order equations.
They are all stable, and this can be shown as follows: the Hamiltonian for BPS states are given by
\be\label{hmq}
H=-\frac{d^2}{dx^2}+W^2_{\phi\phi}+W_\phi\, W_{\phi\phi\phi}
\ee
and now we can introduce the first-order operators
\be
S_{\pm}=-\frac{d}{dx}\pm W_{\phi\phi}
\ee
and they allow introducing the Hamiltonians
\be\label{hfactor}
H_\pm=S^{\dag}_{\pm}S_{\pm}
\ee
which can be written in the general form
\be
H_{\pm}=-\frac{d^2}{dx^2}+W^2_{\phi\phi}\pm W_\phi W_{\phi\phi\phi}
\ee
These Hamiltonians are supersymmetric partners \cite{susyqm1,susyqm2} in quantum mechanics. Now, since $H_{+}$ is the Hamiltonian 
of equation (\ref{hmq}), we see that it is non-negative, thus it has no negative eigenvalue, and this ensure linear stability for the BPS states. We can also see that the zero-mode $\eta_0(x)$ obeys
\be
\frac{d\eta_0(x)}{dx}=\pm W_{\phi\phi}\,\eta_0(x)
\ee
and it can be written as 
\be
\eta_0(x)=A e^{\pm\int dx\,W_{\phi\phi}}
\ee
where $\phi=\phi(x)$ stands for the static solution. The constant $A$ is used to normalize the wave function, but this only
occurs for one of the two signs of the exponential. There is another way to write the zero-mode; it is
\be\label{mz1}
\eta_0=A W_\phi
\ee
where $\phi=\phi(x)$ is the static solution, the BPS state. The zero-mode $\eta_0(x)$ is also known as the bosonic zero-mode,
since the inclusion of fermions in the model will also lead to fermionic zero-modes \cite{jr}.

We consider the $\phi^4$ model. The topological solutions are $\phi_\pm(x)=\pm\tanh(x)$ and the potential
for the corresponding quantum-mechanical problem has the form 
\be\label{pt4}
U(x)=\frac{d^2V}{d\phi^2}\biggr|_{\phi_\pm(x)}=4-6\, {\rm sech}^2(x)
\ee
This is the modified P\"oschl-Teller potential, which was investigated long ago \cite{mfe,flu}. We consider the potential
\be\label{potpt1}
U(x)=a+b\,\tanh(x)-c\;{\rm sech}^2(x)
\ee
with $a,b,c$ real and positive parameters. For $a>b$ and $a<c,$ the continuum spectrum presents reflecting states, for energies
in between $a-b$ e $a+b,$ and other states for energies higher than $a+b.$ If $b$ is zero, the continuum does not contain
reflecting states. For $a>b$ and $a<c,$ the discrete spectrum may contain the zero-mode and other states, with positive and negative eigenvalues. In fact, the bound states present eigenvalues given by  
\be
\epsilon_n=a-A_n^2-\frac{b^2}{4A^2}
\ee
where
\be
A_n=\sqrt{c+\frac14}-n-\frac12
\ee
and $n=0,1,...,$ is such that 
\be\label{potpt4}
n<\sqrt{c+\frac14}-\frac12-\sqrt{\frac{b}{2}}
\ee
In Ref.~\cite{mfe} there is a misprint: the last term of the result for the bound states there obtained has to be multiplied by the factor $1/4.$

The above results can be used to show that for the $\phi^4$ model, the potential (\ref{pt4}) leads to two bound states, one being the zero mode, and the other having eigenvalue $3.$ For the inverted $\phi^4$ model, the potential for the quantum-mechanical problem
is 
\be\label{pt4i}
U(x)=1-6\,{\rm sech}^2(x)
\ee
It has the very same form of the potential for the $\phi^4$ model, but it plots differently, in a way such that 
its zero mode corresponds to the higher bound state, thus introducing a negative eigenvalue, $-3,$ which makes the lump
unstable.

%%%%%%%%%%%%%%%%%%%%%%%%%%%%%%%%%%%%%%%%%%%%%%%
%%%%%%%%%%%%%%%%%%%%%%%%%%%%%%%%%%%%%%%%%%%%%%%
%%%%%%%%%%%%%%%%%%%%%%%%%%%%%%%%%%%%%%%%%%%%%%%
\subsection{Examples}

We now consider some examples, to illustrate the above results. The first example is the sine-Gordon model, which is described by the potential
\be
V(\phi)=\frac12\sin^2(\phi)
\ee
Here we can write $W(\phi)=-\cos(\phi).$ The first-order equations are
\be
\frac{d\phi}{dx}=\pm\sin(\phi)
\ee
The minima of the potential are given by ${\bar\phi}_0=0,$ $\bar\phi_{\pm1}=\pm\pi,$
etc. There is an infinity of topological sectors. They are all equivalent, and in the sectors 
${\bar\phi}_{-1}\leftrightarrow{\bar\phi}_0$ and ${\bar\phi}_0\leftrightarrow{\bar\phi}_1$ the solutions can be written as
\be
\phi^{\pm}_\pm(x)=\pm 2\,{\arctan}(e^{\pm x})
\ee
The energy is $E=E_{BPS}=2.$

Stability investigations lead us to the quantum mechanical potential
\be
U(x)=1-2\,{\rm sech}^2(x)
\ee
which has just one bound state, the zero mode, with normalized eigenfunction given by
\be
\eta_0(x)=\sqrt{\frac12}\, {\rm sech}(x)
\ee
There are other sectors, and there are other solutions; see, e.g., Ref.~\cite{B1,B4}.
 
Another example is the double sine-Gordon model. It may be described by the potential
\be
V(\phi)=\frac2{1+r}(r+\cos(\phi))^2
\ee
where $r$ is a real parameter. This model was recently investigated in \cite{phyD}. It can be described by the superpotential
\be
W(\phi)=\frac2{\sqrt{1+r}}(r\phi+\sin(\phi))
\ee
We consider $r\in(0,1)$ and the singular points, $dW/d\phi=0,$ identify the minima of the potential.
They are periodic and, in the interval $-2\pi<\phi<2\pi,$ we can identify ${\bar\phi}=\pm\pi\pm\arccos(r).$ We see that the maxima
in $\pm\pi$ and the minima in $\pm\pi\pm\arccos(r)$ collapse to two minima in $\pm\pi$ in the limit $r\to1.$ This fact is illustrated
in Fig.~5, where we plot the double sine-Gordon potential for the values $r=1/3,2/3$ and $r=1.$

%%%%%%%%%%%%%%%%%%%%%%%%%%%%%%%%%%%%%%%
\begin{figure}[ht]
\begin{center}
\includegraphics[height=6cm,width=10cm]{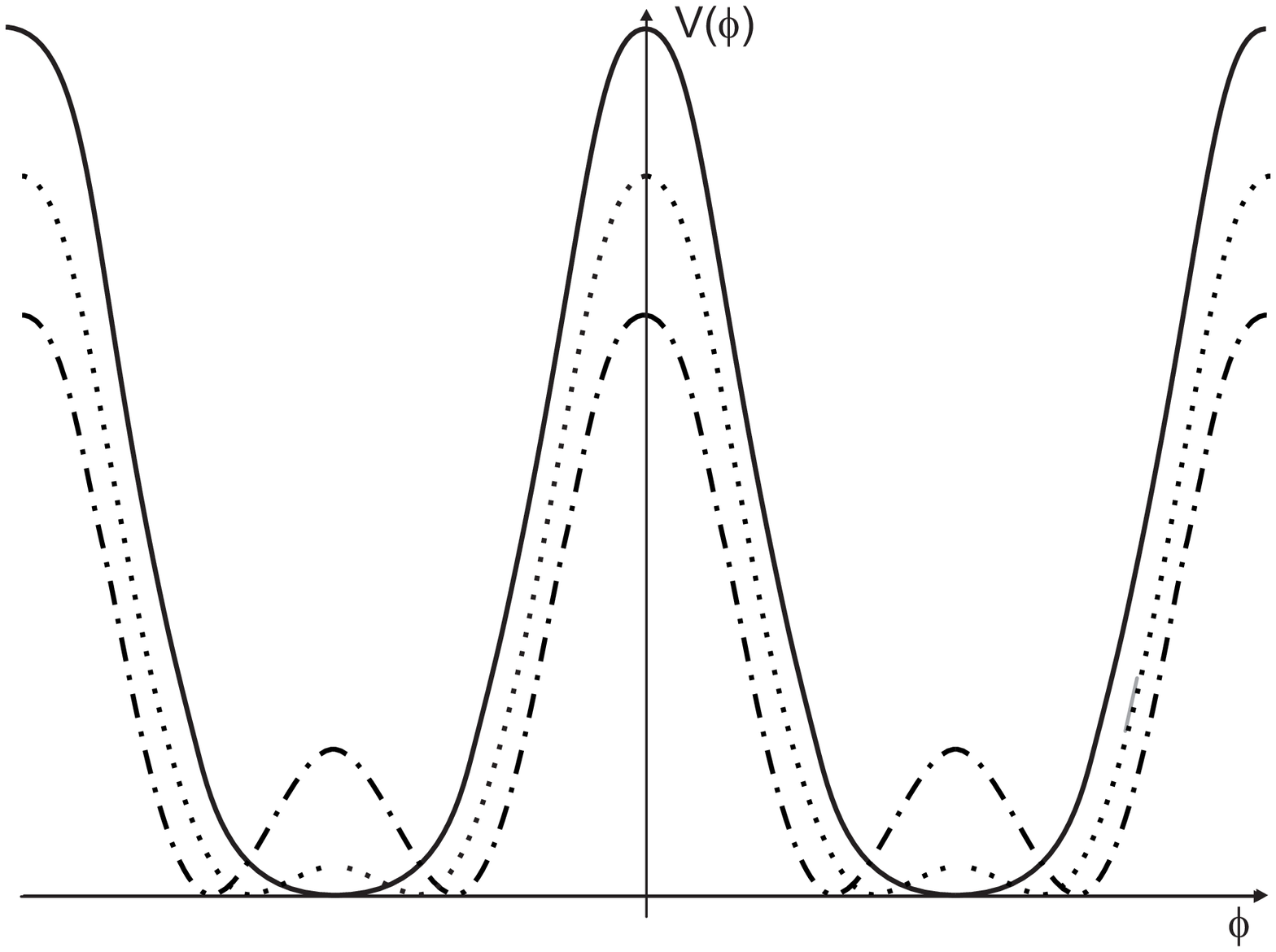}
\end{center}
\caption{Potential for the double sine-Gordon model, for $r=1/3,$ $r=2/3,$ and $r=1,$
depicted with dot-dashed, dashed and solid lines, respectively.}
\end{figure}
%%%%%%%%%%%%%%%%%%%%%%%%%%%%%%%%%%%%%%%%%%%%%%%%%%%%

This model contains two topological sectors, one large and the other small, with energies
\be
E^l_{BPS}=4\sqrt{1-r}+4r\frac{\pi-\arccos(r)}{\sqrt{1+r}}
\ee
and
\be
E^s_{BPS}=4\sqrt{1-r}-4r\frac{\arccos(r)}{\sqrt{1+r}}
\ee
We notice that $E^l_{BPS}=E^s_{BPS}+ 4\pi r/\sqrt{1+r},$ and in the limit $r\to1$ we get $E^l_{PBS}\to 2\sqrt{2}\pi$ e $E^s_{BPS}\to0.$

Stable solutions which solve the first-order equations are given by
\be
\phi^g_\pm(x)=\pm2 \arctan\biggl[\sqrt\frac{1+r}{1-r}\tanh(\sqrt{1-r}x)\biggr]
\ee 
and
\be
\phi^{p}_\pm(x)=\pm\pi-2 \arctan\biggl[\sqrt\frac{1+r}{1-r}\tanh(\sqrt{1-r}x)\biggr]
\ee

It is interesting to see that the double sine-Gordon model can be used to map specific magnetic materials, like the one studied in Ref.~\cite{brw}; there one identifies three distinct phases: anti-ferromagnetic, canted, and weakly ferromagnetic.
These phases can be described by the above model, in the cases $r=0,$ $0<r<1$ and $r=1,$ respectively. Moreover,
in the canted phase, there also appear two distinct defect structures, the large and small solutions. 

Another example is described by the $\phi^6$ model. Here the potential has the form
\be
V(\phi)=\frac12\phi^2(1-\phi^2)^2
\ee
We have $W(\phi)=(1/2)\phi^2-(1/4)\phi^4,$ and the minima are $\bar\phi_0=0$ and $\bar\phi_\pm=\pm1.$ There are two topological sectors, equivalent. The solutions are 
\be
\phi^\pm_\pm(x)=\pm\,\sqrt{\frac12[1\pm\tanh(x)]}
\ee
and they identify the kinks and anti-kinks in the two sectors. Their energies are given by $E=E_{BPS}=1/4.$

Stability leads us to the quantum-mechanical potentials
\be
U_\pm(x)=\frac{5}{2}\mp\frac{3}{2}\tanh(x)-\frac{15}{4}\,{\rm sech}^2(x)
\ee
The terms which contain the hyperbolic tangent make the potentials asymmetric. This introduces reflection states into the continuum,
although there is only one bound state, the zero-mode; see Ref.~\cite{lohe} for more details.

Another model is given by \cite{hkd}
\be\label{pp4}
V(\phi)=\frac12(1-|\phi|)^2
\ee
It is similar to the $\phi^4$ model. It has minima in $\bar\phi_\pm=\pm1,$ and we can write $W(\phi)=\phi-(1/2)|\phi|\phi.$ The first-order equations are
\be
\frac{d\phi}{dx}=\pm(1-|\phi|)
\ee 
The solutions are 
\be\label{pp4s}
\phi_\pm(x)=\pm 2\,\frac{\tanh(x/2)}{1+\tanh(|x|/2)}
\ee
Their energies are given by $E=E_{BPS}=1.$ In Figs.~6 and 7 we plot potentials and kinks, for the above model, and for the $\phi^4$
model, for comparison.

%%%%%%%%%%%%%%%%%%%%%%%%%%%%%%%%%%%%%%%%%%%%%%%%%%%
\begin{figure}[ht]
\begin{center}
\includegraphics[height=6cm,width=10cm]{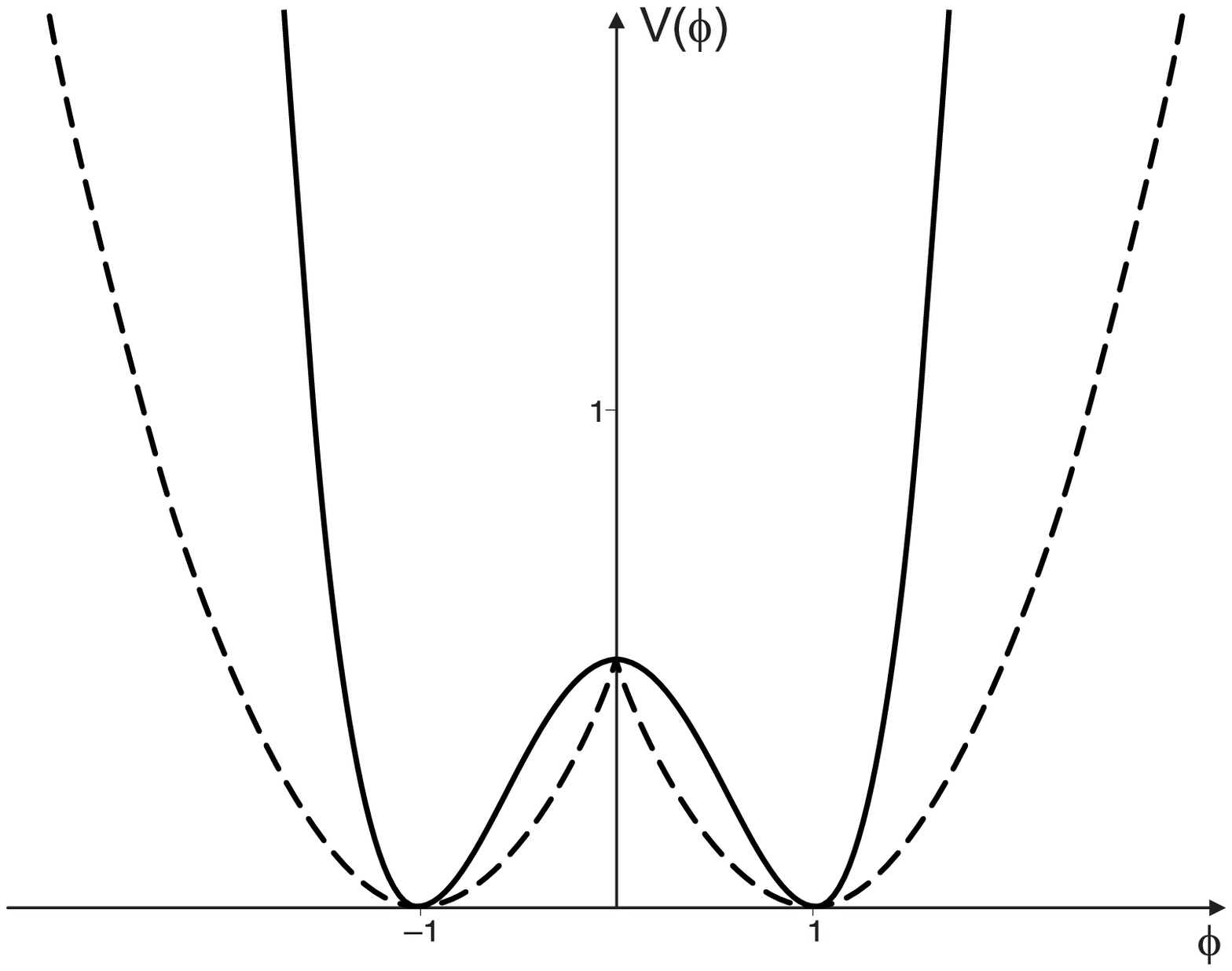}
\end{center}
\caption{The potential of Eq.~(\ref{pp4}), depicted with dashed line, and the potential of the $\phi^4$ model.}
\end{figure}
%%%%%%%%%%%%%%%%%%%%%%%%%%%%%%%%%%%%%%%%%%%%%%%%%%%
\vspace{1cm}
%%%%%%%%%%%%%%%%%%%%%%%%%%%%%%%%%%%%%%%%%%%%%%%%%%%
\begin{figure}[ht]
\begin{center}
\includegraphics[height=6cm,width=10cm]{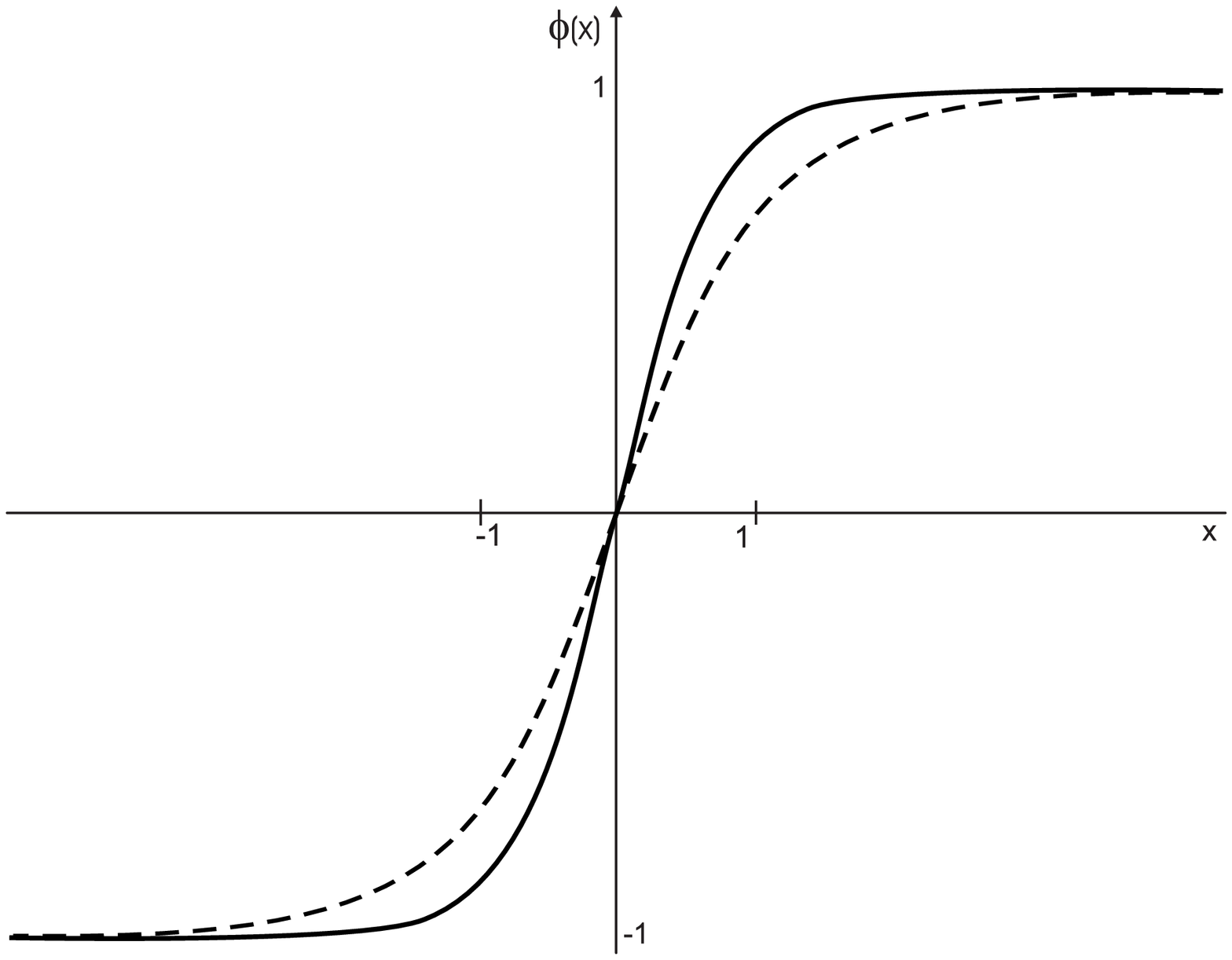}
\end{center}
\caption{The kink of Eq.~(\ref{pp4s}), depicted with dashed line, and the kink of the $\phi^4$ model.}
\end{figure}
%%%%%%%%%%%%%%%%%%%%%%%%%%%%%%%%%%%%%%%%%%%%%%%%%%%

Stability investigations lead us to the potential
\be
U(x)=1-2\delta(x)
\ee
which is attractive. There is just one bound state, the zero mode. To see this explicitly, note that the Hamiltonian
can be written as
\be
-\frac{d^2}{dx^2}+1-2\,\delta(x)=
\left(-\frac{d}{dx}+\frac{|x|}{x}\right)\left(\frac{d}{dx}+\frac{|x|}{x}\right)
\ee 
Thus, we see that the zero-mode obeys 
\be
\left(\frac{d}{dx}+\frac{|x|}{x}\right)\eta_0(x)=0
\ee
This equation can be solved easily, given the normalized zero-mode
\be
\eta_0(x)=\exp(-|x|)
\ee

The above model was investigated in \cite{hkd,td}, leading to exact solutions for a model described by coupled linear chains.
More recently, it was studied in \cite{th,bil} with other motivations. In particular, in \cite{bil} we have introduced another model, inspired in the above model and in the $\phi^6$ model. It is described by
\be\label{nm1}
V(\phi)=\frac12\phi^2(1-|\phi|)^2
\ee
There are minima in $\bar\phi=0$ and $\bar\phi_\pm=\pm1,$ and we have $W(\phi)=(1/2)\phi^2-(1/3)|\phi|\phi^2.$ There are two topological sectors, equivalent, with energies given by $E=E_{BPS}=1/6.$ The solutions are
\ben\label{pp6s}
\phi^+_\pm(x)=\pm\frac12[1+\tanh(x/2)]
\\
\phi^-_\pm(x)=\pm\frac12[1-\tanh(x/2)]
\een
They are similar to the solutions of the $\phi^6$ model; see Figs.~8 and 9 for a comparison.

%%%%%%%%%%%%%%%%%%%%%%%%%
\begin{figure}
\begin{center}
\includegraphics[height=6cm,width=10cm]{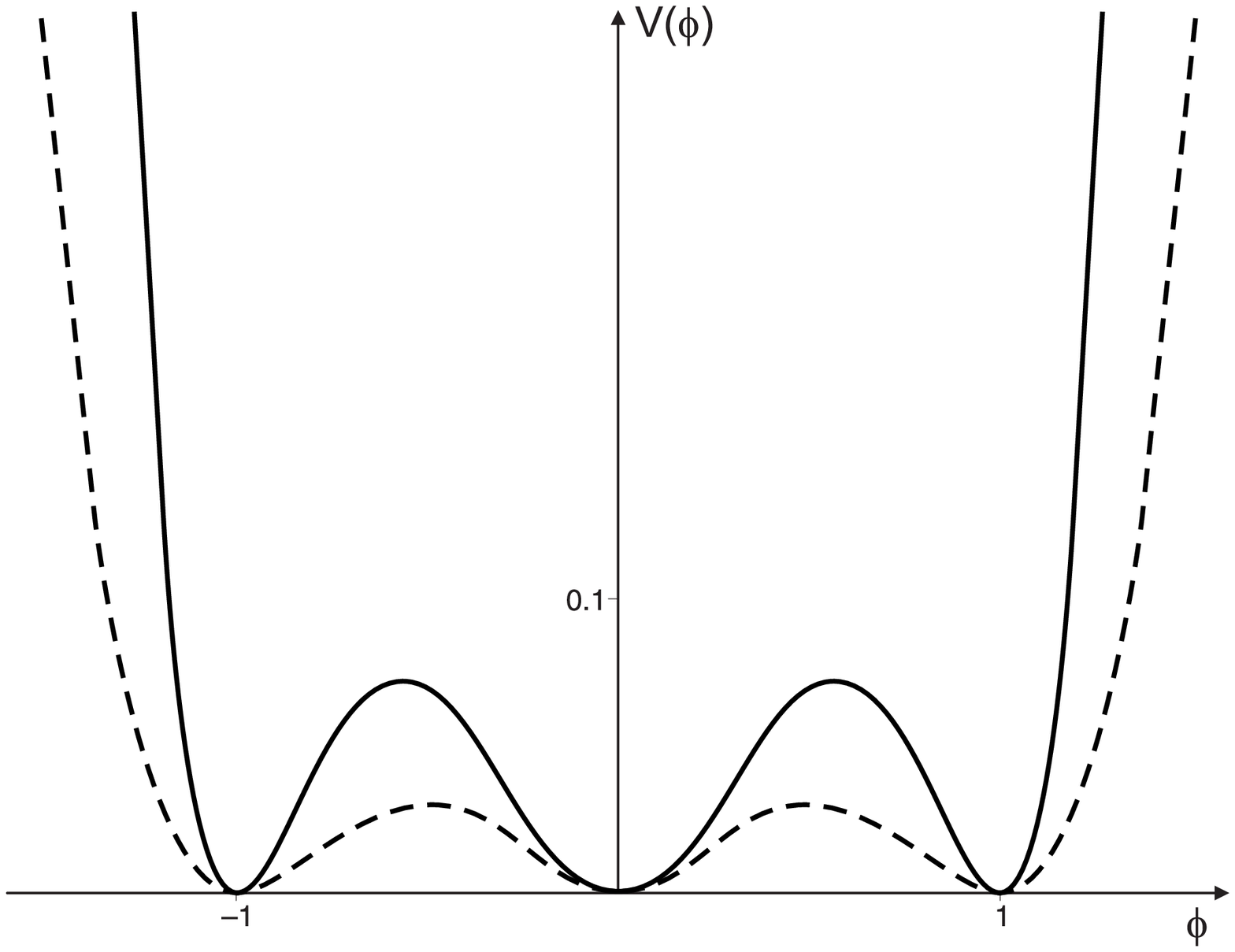}
\end{center}
\caption{The potential of the Eq.~(\ref{nm1}), depicted with dashed line, and the potential of the $\phi^6$ model.}
\end{figure}
%%%%%%%%%%%%%%%%%%%%%%%%%

%%%%%%%%%%%%%%%%%%%%%%%%%%%%%%%%%%
\begin{figure}
\begin{center}
\includegraphics[height=6cm,width=10cm]{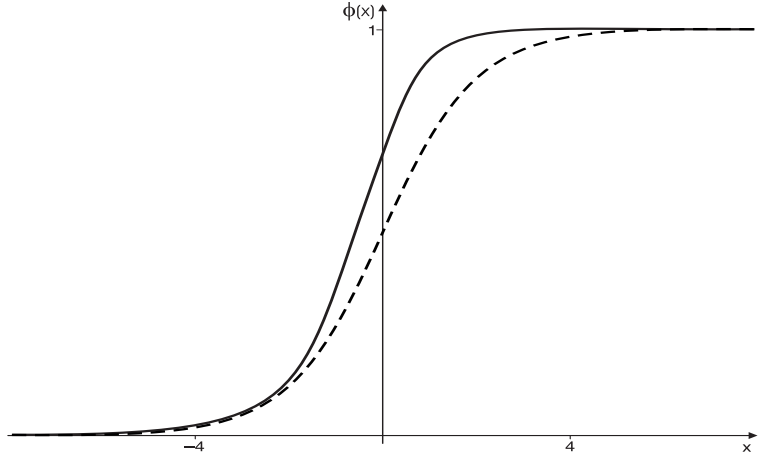}
\end{center}
\caption{The kink of the Eq.~(\ref{pp6s}), depicted with dashed line, and the kink of the $\phi^6$ model.}
\end{figure}
%%%%%%%%%%%%%%%%%%%%%%%%%%%%%%%%%%

An interesting feature of this new model is that the topological sectors are symmetric, related to the local maxima they have.
This makes the topological solutions uniform, leading to a symmetric quantum-mechanical potential which contains no reflection states,
as it appeared in the $\phi^6$ model. This model is of direct interest to study first-order phase transition, and we believe that it may be of some use in applications to Condensed Matter.

We investigate stability of the solutions of the above model. We get to the potential
\be\label{stap}
U(x)=1-\frac32{\rm sech}^2(x/2)
\ee
The Hamiltonian can be factorized as 
\be
\left(-\frac{d}{dx}+\tanh(x/2)\right)\left(\frac{d}{dx}+\tanh(x/2)\right)
\ee
Thus, the zero-mode should obey 
\be
\left(\frac{d}{dx}+\tanh(x/2)\right)\eta_0(x)=0
\ee
We integrate this equation to obtain the normalized zero-mode
\be
\eta_0(x)=\frac12\sqrt{\frac32}{\rm sech}^2(x/2)
\ee
There is another bound state, with eigenvalue $3/4,$ and no reflection state appears in the continuum.

Another example is given by the potential \cite{cv1}
\be\label{nm2}
V(\phi)=\frac12 {\rm sech}^2(\phi)
\ee
The novelty here is that the potential has no minimum, but it supports topological defects, as shown in Refs.~\cite{cv1,cv2}.
Our participation in this problem started with \cite{b99}, where we have shown that this model engenders first-order equations
given in terms of the superpotential
\be
W(\phi)={\rm arctan}[\sinh(\phi)]
\ee
The first-order equations are
\be
\frac{d\phi}{dx}=\pm\,{\rm sech(\phi)}
\ee
They are solved by 
\be
\phi_\pm(x)=\pm {\rm arcsinh}(x)
\ee
These defects are different: they are very diffuse, as we see from the energy density
\be
\varepsilon(x)=\frac1{1+x^2}
\ee
However, the energy is finite, given by $E=E_{BPS}=\pi.$   

It is interesting to see that if we write the topological current in the usual form $j^\mu=(1/2)\varepsilon^{\mu\nu}\partial_\nu\phi,$ we would have problem to calculate the topological charge. However, since the topological current is not unique, we can write
\be
j^\mu=\varepsilon^{\mu\nu}\partial_\nu W(\phi)
\ee 
In this case, we are relating the topological charge with the energy of the solution, which seems to be appropriate, since the energy
should be well-defined.

We investigate stability of the above solutions to get to
\be
U(x)=\frac{-1+2x^2}{(1+x^2)^2}
\ee
This is a volcano-like potential, as we show in Fig.~10. It has a single bound state, the zero-mode, with normalized eigenfunction
given by
\be
\eta_0(x)=\sqrt{\frac1{\pi}}\sqrt{\frac1{1+x^2}}
\ee
In this case, there is no gap between the zero-mode and the continuum. 

%%%%%%%%%%%%%%%%%%%%%%%%
\begin{figure}
\begin{center}
\includegraphics[height=6cm,width=10cm]{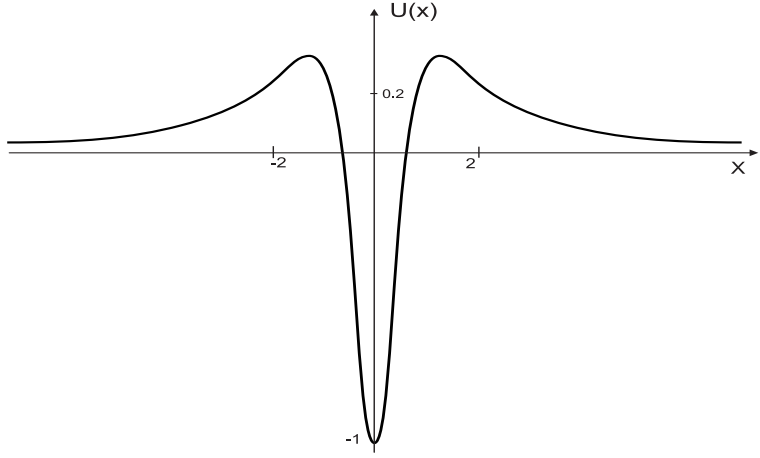}
\end{center}
\caption{The volcano-like potential.}
\end{figure}
%%%%%%%%%%%%%%%%%%%%%%%%%%%%%%%%

Another model is described by the potential \cite{blm1}
\be\label{nm3}
V(\phi)=\frac12(1-\phi^2)^3
\ee
The novelty here is that this potential is unbounded from below, and it has two inflection points, at the values $\bar\phi=\pm1,$ such that $V(\bar\phi)=0.$ However, there are topological solutions connecting these inflection points. They are given by
\be
\phi_\pm(x)=\pm\frac{x}{\sqrt{1+x^2}}
\ee
The appearance of topological solutions connecting inflection points was first shown in \cite{blm1}.

We investigate stability to get to the potential 
\be
U(x)=12\frac{-1/4+x^2}{(1+x^2)^2}
\ee
which is also volcano-like. It has only one bound state, the zero-mode, and there is no gap between the zero-mode and the continuum.
The zero-mode is given by
\be
\eta_0(x)=2\sqrt{\frac{2}{3\pi}}\sqrt{\frac1{(1+x^2)^3}}
\ee 

In the last two examples, there is no gap between the zero-mode and the continuum, and so we should investigate how the quantum
corrections should appear, to see if there are subtleties, as it occurs in models where a gap is present; see, e.g., 
Refs~\cite{nie,jaf,shi1} for more information on this issue.

Another model is described by the potential \cite{bmm}
\be\label{nm4}
V(\phi)=\frac12(\phi^{(p-1)/p}-\phi^{(p+1)/p})^2
\ee
where $p$ is a real parameter. For simplicity, here we consider $p$ odd integer, $p=1,3,5,...$ This model can be seen as a generalization of the $\phi^4$ model, which is obtained for $p=1.$ 

In this case we can write
\be
W(\phi)=\frac{p}{2p-1}\phi^{(2p-1)/p}-\frac{p}{2p+1}\phi^{(2p+1)/p}
\ee
and the first-order equations are solved by 
\be
\phi_\pm(x)=\pm\tanh^p(x/p)
\ee
The energies are given by $E^p_{BPS}=4p/(4p^2-1)$. In Fig.~11 we plot some solutions, and there we see that they connect the minima
$\pm1,$ passing through the minimum at zero with vanishing derivative.

%%%%%%%%%%%%%%%%%%%%%%%%
\begin{figure}[ht]
\begin{center}
\includegraphics[height=6cm,width=10cm]{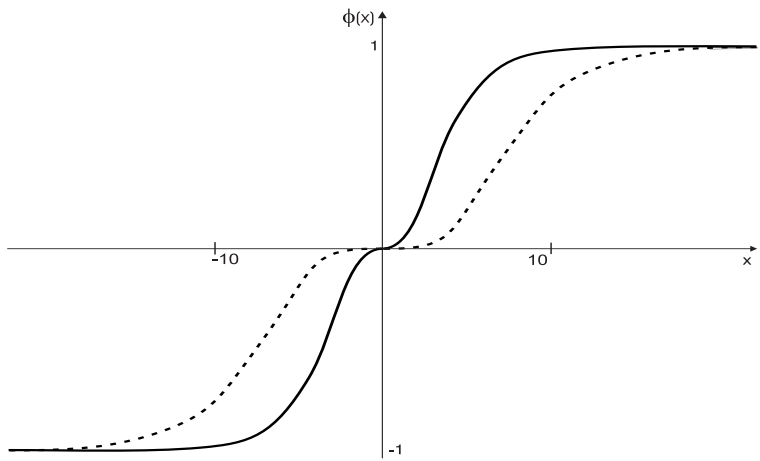}
\end{center}
\caption{The 2-kink solution for $p=3$ and $5,$ depicted with solid and dashed lines, respectively.}
\end{figure}
%%%%%%%%%%%%%%%%%%%%%%%%%%%%%%%%

As far as we can see, this is the first time that topological solutions connect non-adjacent minima, passing through an intermediate minimum. This fact is only possible because the minimum at zero is not a true minimum, that is, it is not good for quantization because it cannot be used as a perturbative ground state, since the mass squared at $\bar\phi=0$ is divergent. For this reason, the above solutions are named 2-kink solutions.

The above solutions present normalized zero-mode, described by
\be
\eta_0(x)=\sqrt{\frac{4p^2-1}{4p}}\tanh^{(p-1)}\left(\frac{x}{p}\right)
{\rm sech}^2\left(\frac{x}{p}\right)
\ee 
They have maxima at the points 
\be
x_\pm=\pm p\,{\rm arctanh}\sqrt{\frac{p-1}{p+1}}
\ee
which can be used to identify how the two kinks localize inside the 2-kink solution.

%%%%%%%%%%%%%%%%%%%%%%%%%%%%%%%%%%%%%%%%%%%%%%%%%%%%%%%%%%%
%%%%%%%%%%%%%%%%%%%%%%%%%%%%%%%%%%%%%%%%%%%%%%%%%%%%%%%%%%%
%%%%%%%%%%%%%%%%%%%%%%%%%%%%%%%%%%%%%%%%%%%%%%%%%%%%%%%%%%%
\section{Two-field models}
\bigskip

Let us now consider models described by two real scalar fields, given by
\be
{\mathcal L}=\frac12\partial_\mu\phi\partial^\mu\phi+
\frac12\partial_\mu\chi\partial^\mu\chi-V(\phi,\chi)
\ee
where $V(\phi,\chi)$ is the potential, which specifies the model to be studied. In $(1,1)$ dimensions, the equations of motion are
\ben\label{em21}
&&\frac{\partial^2\phi}{\partial t^2}-\frac{\partial^2\phi}{\partial x^2}+
\frac{\partial V}{\partial\phi}=0
\\\label{em22}
&&\frac{\partial^2\chi}{\partial t^2}-\frac{\partial^2\chi}{\partial x^2}+
\frac{\partial V}{\partial\chi}=0
\een
and for static solutions we have
\ben\label{ems21}
&&\frac{d^2\phi}{dx^2}=\frac{\partial V}{\partial\phi}
\\\label{ems22}
&&\frac{d^2\chi}{dx^2}=\frac{\partial V}{\partial\chi}
\een

Non-linearity makes it hard to deal with these equations on general grounds. Some investigations were done in Refs.~\cite{rw,mon,ruck}. In particular, in \cite{rw} one illustrates the difficulties to solve Eqs.~(\ref{ems21}) and (\ref{ems22}). This problem was further  considered in \cite{B179} and there it was proposed a method, named trial orbit method, which could perhaps be of some use
to solve the above equations. In 1995 we started our investigations on models described by two real scalar fields.
We started with the potential \cite{bds}
\be\label{pot2}
V(\phi,\chi)=\frac12 W^2_\phi+\frac12 W^2_\chi
\ee
In this case, the equations of motion for static fields can be written in the form
\ben\label{ems121}
&&\frac{d^2\phi}{dx^2}=W_\phi\,W_{\phi\phi}+W_\chi\,W_{\chi\phi}
\\\label{ems122}
&&\frac{d^2\chi}{dx^2}=W_\phi\,W_{\phi\chi}+W_\chi\,W_{\chi\chi}
\een
and the energy associated with these configurations could be written as
\be
E=\frac12\int^{\infty}_{-\infty} dx
\biggl[\left(\frac{d\phi}{dx}\right)^2+\left(\frac{d\chi}{dx}\right)^2+
W^2_\phi+W_\chi^2\biggr]
\ee
This expression can be rewritten in the form
\be
E=E_B+\frac12\int^{\infty}_{-\infty} dx
\biggl[\left(\frac{d\phi}{dx}\mp W_\phi\right)^2+
\left(\frac{d\chi}{dx}\mp W_\chi\right)^2\biggr]
\ee
where we have set
\be
E_B=E_{BPS}=|\Delta W|
\ee
with
\be
\Delta W= W(\phi(\infty),\chi(\infty))-W(\phi(-\infty),\chi(-\infty))
\ee
This procedure shows that the energy is minimized to $E=E_{BPS},$ for field configurations which obey the first-order equations
\ben\label{eb1}
\frac{d\phi}{dx}=\pm W_\phi
\\\label{eb2}
\frac{d\chi}{dx}=\pm W_\chi
\een
On the other hand, if we differentiate these equations with respect to $x,$ we can see that their solutions also solve the second-order equations of motion for $W_{\phi\chi}=W_{\chi\phi},$ which is the condition for $W$ to present second derivative, as the potential (\ref{pot2}) asks implicitly.

It is interesting to see that solutions of the first-order equations have the same portion of gradient and potential energies.
The first-order equations can be seen as a dynamical system, and so we can use all the mathematical tools available for dynamical systems to deal with these equations. In particular, the algebraic equations $W_\phi=0$ and $W_\chi=0$ can be solved to give
the set of singular points of the respective dynamical system. In the field theoretical model, this set of singular points constitute the absolute minima of the model. We can characterize the absolute minima with $v_i=(\phi_i,\chi_i), i=1,2,...$ Thus, each pair of minima $v_i$ e $v_j$ may represent a topological sector, which may be BPS or non-BPS.

A given sector $ij$ is BPS if the energy $E^{(ij)}$ is not zero, that is, if $W_i=W(\phi_i,\chi_i)$ is distinct from $W_j=W(\phi_j,\chi_j),$ to make the energy $E^{(ij)}_{BPS}=|\Delta W_{ij}|=|W_i-W_j|$ non-vanishing. On the other hand,
if $W_i=W(\phi_i,\chi_i)$ is equal to $W_j=W(\phi_j,\chi_j),$ there is no BPS energy available, and this shows that there is no BPS solution connecting the minima that identify the sector. In this case, however, the sector may have configurations which solve the equations of motion. These sectors are usually named non-BPS, because the field configurations do not solve the first-order equations.

In the BPS limit, models described by two real scalar fields are simpler, because we have to deal with pairs of first-order equations.
We then focus on the equations (\ref{eb1}) and (\ref{eb2}). We can write 
\be
W_\chi\, d\phi-W_\phi\,d\chi=0
\ee
This equation is easily solved when $W(\phi,\chi)$ is harmonic. In this case, the solutions describe an orbit in the $(\phi,\chi)$
plane, given by $F(\phi,\chi)=0,$ which is obtained with
\be
\frac{\partial F}{\partial\phi}=W_\chi,\,\,\,\,\,
\frac{\partial F}{\partial\chi}=-W_\phi
\ee
In Refs.~\cite{bms1,bms2} we have shown that when $W$ is harmonic, all the solutions are BPS solutions.
 
When $W(\phi,\chi)$ is not harmonic, we have to search for the integrating factor $I(\phi,\chi)$. This factor is needed to make another function, ${\wt F}(\phi,\chi),$ harmonic, such that
\be
\frac{\partial{\wt F}}{\partial\phi}=I(\phi,\chi)\,W_\chi,\,\,\,\,\,
\frac{\partial{\wt F}}{\partial\chi}=-I(\phi,\chi)\,W_\phi
\ee
The main problem with the integrating factor is that we do not know how to get it in general, and so the trial orbit method 
is yet of good use to find solutions in specific models.

The trial orbit method is very effective when it is applied to first-order equations. This was shown in Ref.~\cite{wilson}, and the main steps can be summarized as follows:

{\bf First step.} {\it Choose a BPS sector.} This is done with two minima, $v_i=(\phi_i,\chi_i)$ and $v_j=(\phi_j,\chi_j),$ such that
$W_i=W(\phi_i,\chi_i)\neq W_j=W(\phi_j,\chi_j).$

{\bf Second step.} {\it Choose an orbit.} This is done with a given function $F_{ij}(\phi,\chi)=0,$ which is compatible with the minima that specify the sector, which satisfies $F_{ij}(\phi_i,\chi_i)=0$ and $F_{ij}(\phi_j,\chi_j)=0$.

{\bf Third step.} {\it Try the chosen orbit.} This is done taking the derivative of the chosen orbit to get
\be
\frac{\partial F_{ij}}{\partial\phi}W_\phi+
\frac{\partial F_{ij}}{\partial\chi}W_\chi=0
\ee
If this new expression is compatible with the chosen orbit, it is a good orbit and can be used to decouple the first-order equations.

These three steps are very efficient to solve the first-order equations, to find BPS solutions, as we illustrate below.

%%%%%%%%%%%%%%%%%%%%%%%%%%%%%%%%%%%%%%%%%%%%%%%%%%%%%%%%%%
%%%%%%%%%%%%%%%%%%%%%%%%%%%%%%%%%%%%%%%%%%%%%%%%%%%%%%%%%%
%%%%%%%%%%%%%%%%%%%%%%%%%%%%%%%%%%%%%%%%%%%%%%%%%%%%%%%%%%
\subsection{Linear Stability}

We can investigate linear stability of the BPS solutions in a general way. We write $\phi=\phi(x,t)$ and $\chi=\chi(x,t)$ as
\ben
\phi(x,t)&=&\phi(x)+\eta(x,t)
\\
\chi(x,t)&=&\chi(x)+\xi(x,t)
\een
where $\eta(x,t)$ and $\xi(x,t)$ are fluctuations around the static configurations $\phi(x)$ and $\chi(x),$ which solve the first-order equations. Since the classical solutions are static, we can write
\begin{eqnarray}
\left(
\begin{array}{c}
\eta(x,t)\\{\xi(x,t)}\end{array}\right)=
\sum_n 
\left(
\begin{array}{c}
\eta_n(x)\\
{\xi_n(x)}\end{array}\right)
\cos(w_n t)
\end{eqnarray}
We use these expressions into the equations of motion (\ref{em21}) and (\ref{em22}) to get to the Schr\"odinger-like equation $H\Psi_n(x)=w^2_n\Psi_n(x),$ where 
\begin{eqnarray}\label{mq2}
H=-\frac{d^2}{dx^2}+
\left(
\begin{array}{cc}
V_{\phi\phi}&V_{\phi\chi}\\
V_{\chi\phi}&V_{\chi\chi}
\end{array}
\right)
\end{eqnarray}
and
\ben
\Psi_n(x)=
\left(
\begin{array}{c}
\eta_n(x)\\
{\xi_n(x)}\end{array}\right)
\een

We consider the potential (\ref{pot2}) to get
\ben
V_{\phi\phi}&=&W^2_{\phi\phi}+W_\phi W_{\phi\phi\phi}+
W^2_{\chi\phi}+W_\chi W_{\chi\phi\phi}
\\
V_{\chi\chi}&=&W^2_{\phi\chi}+W_\phi W_{\phi\chi\chi}+
W^2_{\chi\chi}+W_\chi W_{\chi\chi\chi}
\een
and
\be
V_{\phi\chi}=V_{\chi\phi}=W_{\phi\chi}W_{\phi\phi}+W_\phi W_{\phi\phi\chi}+
W^2_{\chi\chi}W_{\chi\phi}+W_\chi W_{\chi\phi\chi}
\ee
In Ref.~\cite{bs} we have shown how to introduce first-order operators such that
\be
S_{\pm}=-\frac{d}{dx}\pm
\left(
\begin{array}{cc}
W_{\phi\phi}&W_{\phi\chi}\\
W_{\chi\phi}&W_{\chi\chi}
\end{array}
\right)
\ee
which lead us with the Hamiltonians
\be
H_{\pm}=S^{\dag}_{\pm}\,S_{\pm}
\ee
Now, in (\ref{mq2}) we see that $H$ is just $H_+,$ the partner of $H_-.$ The Hamiltonians $H_+$ e $H_-$ can be seen as supersymmetric partners in quantum mechanics. The factorization of $H$ implies that it is non-negative, and this proves stability of the BPS solutions in general. Moreover, the zero-mode solves the equation 
\be
\frac{d\Psi_0}{dx}=
\left(
\begin{array}{cc}
W_{\phi\phi}&W_{\phi\chi}\\
W_{\chi\phi}&W_{\chi\chi}
\end{array}\right)\Psi_0
\ee
and so it has the form 
\be\label{mz2}
\Psi_0(x)=A
\left(
\begin{array}{c}
\eta_0(x)\\
\xi_0(x)
\end{array}\right)=A
\left(
\begin{array}{c}
W_\phi\\W_\chi
\end{array}\right)
\ee
where $A$ is the normalization constant. This result can be seen as a natural extension of the former result, for a single field, given by Eq.~(\ref{mz1}).

%%%%%%%%%%%%%%%%%%%%%%%%%%%%%%%%%%%%%%%%%%%%%%%%%%%%
%%%%%%%%%%%%%%%%%%%%%%%%%%%%%%%%%%%%%%%%%%%%%%%%%%%%
%%%%%%%%%%%%%%%%%%%%%%%%%%%%%%%%%%%%%%%%%%%%%%%%%%%%
\subsection{Example}

We illustrate the above investigations with an example. We consider the function
\be
W(\phi,\chi)=\phi-\frac13\,\phi^3-r\,\phi\,\chi^2
\ee
where $r$ is a real parameter. The potential has the form 
\be\label{bnrt}
V(\phi,\chi)=\frac12(1-\phi^2)^2-r\chi^2+r(1+2r)\phi^2\chi^2+\frac12r^2\chi^4
\ee
This model was first studied in Ref.~\cite{bds}. See also Refs.~\cite{brs1,bnrt,csh,svo,spa1,spa2,spa3,bb1,bb2,etb,bbb}
for related investigations.

The potential (\ref{bnrt}) is plotted in Fig.~12. For static solutions the equations of motion are given by
\ben
\frac{d^2\phi}{dx^2}&=&-2\phi+2r(1+2r)\phi\chi^2+2\phi^3
\\
\frac{d^2\chi}{dx^2}&=&-2r\chi+2r(1+2r)\phi^2\chi+2r^2\chi^3
\een
The first-order equations are much simpler
\ben
\frac{d\phi}{dx}&=&\pm1\mp\phi^2\mp r\chi^2
\\
\frac{d\chi}{dx}&=&\mp2r\phi\chi
\een
The minima of the potential are $(\pm1,0)$ for $r<0$, or $(\pm1,0)$ and $(0,\pm1/\sqrt{r})$ for $r>0.$ We consider the richer case, for $r>0.$ Here there are four minima, given by $v_1=(1,0),\,v_2=(-1,0),\,v_3=(0,1/\sqrt{r})$ and $v_4=(0,-1/\sqrt{r}).$ There are six
topological sectors: sector 1, which connects the minima $v_1$ and $v_2;$ sector 2, which connect the minima $v_3$ and $v_4;$ 
sector 3, which connect the minima $v_1$ and $v_3;$  sector 4, which connect the minima $v_1$ and $v_4;$ sector 5, which connect the minima $v_2$ and $v_3;$ and, finally, sector 6, which connect the minima $v_2$ and $v_4.$

%%%%%%%%%%%%%%%%%%%%%%%%%%%%%%%%%%%
\begin{figure}[ht]
\begin{center}
\includegraphics[angle=90,width=10cm]{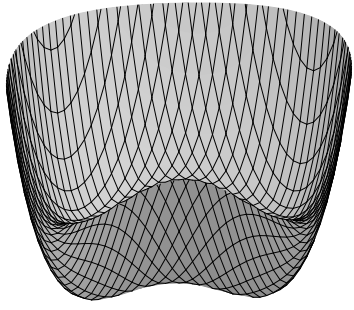}
\end{center}
\caption{Potential for the two-field model of Eq.~(\ref{bnrt}).}
\end{figure}
%%%%%%%%%%%%%%%%%%%%%%%%%%%%%%%%%%%

The value of $W$ at the minima are given by: $W_1=W(1,0)=2/3,$ $W_2=W(-1,0)=-2/3,$ $W_3=W(0,1/\sqrt{r})=0,$ and $W_3=W(0,-1/\sqrt{r})=0.$ Thus, there are five sectors which are BPS, one with energy $E^1_{BPS}=4/3$ and the other four with the same energy $E^3_{BPS}=E^4_{BPS}=E^5_{BPS}=E^6_{BPS}=2/3.$ There is one sector which is non-BPS; it has energy $E^2_{nBPS},$
which can only be obtained when one shows the corresponding topological solution explicitly.

To find explicit solutions, let us first consider solutions which depend only on a single field. We first set $\chi=0$ in the equations of motion. We get 
\be
\frac{d^2\phi}{dx^2}=-2\phi(1-\phi^2)
\ee
which is the equation for the $\phi^4$ model, which has the solutions $\phi_\pm(x)=\pm\tanh(x),$ with energy $4/3.$ Thus, the pair of solutions 
\be
\phi_\pm(x)=\pm\tanh(x),\,\,\,\,\,\,\chi=0
\ee
are solutions for the above two-field model. In field space, these solutions describe a straight line connecting the minima
$v_1$ and $v_2,$  which identify the topological sector 1. We also see that the imposition $\chi=0$ reduces the first-order
equations to the equation 
\be
\frac{d\phi}{dx}=\pm(1-\phi^2)
\ee
which have the same solutions, showing that they are BPS solutions.

On the other hand, if we set $\phi=0$ in the equations of motion, we get to
\be
\frac{d^2\chi}{dx^2}=-2r^2\chi(1/r-\chi^2)
\ee
which is solved by
\be
\chi_\pm(x)= \pm\frac1{\sqrt{r}}\tanh(\sqrt{r}x)
\ee
with energy $E^2_{nBPS}=4/3\sqrt{r},$ which depend on $r.$ Thus, the pair of solutions
\be
\phi=0,\,\,\,\,\,\,\chi_\pm(x)=\pm\frac1{\sqrt{r}}\tanh(\sqrt{r}x)
\ee
represent straight line orbit which connects the minima $v_3$ and $v_4$. In this case, however, the solutions are non-BPS, because they do not satisfy the first-order equations. We notice that the energy of the non-BPS solutions depend on $r;$ for $r\in(0,1)$ it is higher than the energy of the BPS states in sector $1.$ 

The above model has an interesting feature. For $r=1,$ we can rotate the $(\phi,\chi)$ plane to $\phi_1=\phi+\chi$ and $\phi_2=\phi-\chi,$ to see that $\phi_1$ and $\phi_2$ decouple, describing two single field models, instead of a single model of two fields. See Ref.~\cite{bbb} for other details on the issue.

In order to illustrate the trial orbit method, let us investigate solutions in the several BPS sectors of the model. First, we consider sector $1,$ identified by the minima $v_1=(1,0)$ and $v_2=(-1,0).$ We consider an elliptic orbit, described by $a\phi^2+b\chi^2=c,$ with $a,\,b,\,c$ being real parameters which identify the orbit. This orbit must contain the two minima which identify the sector $1;$ these conditions eliminates two parameters, and we get to the more specific orbit $\phi^2+b\chi^2=1.$ We now derivate this orbit and use the first-order equations to get to $\phi^2+r(1+2b)\chi^2=1,$ which is compatible with our orbit for $b=r/(1-2r).$ Thus, the orbit 
\be
\phi^2+\frac1{\frac1r-2}\,\chi^2=1
\ee
is a good orbit for $r\in(0,1/2),$ and can be used to solve the problem. We use this orbit to write 
\be
\chi^2=\left(\frac1r-2\right)(1-\phi^2)
\ee 
We use this into the first-order equation for $\phi$ to obtain
\be
\frac{d\phi}{dx}=\pm\, 2r\,(1-\phi^2)
\ee
which are solved by 
\be
\phi_\pm(x)=\pm\,\tanh(2rx)
\ee
Thus, for the other field, $\chi,$ we get
\be
\chi_\pm(x)=\pm\sqrt{\frac1r-2\,}\, {\rm sech}(2rx)
\ee
There are four solutions for the pair $(\phi,\chi),$ and in Fig.~13 we plot a pair of solutions for $r=1/6.$ It is interesting to notice that the above solutions can be used to describe solitons in ferroelectric materials, as we have pointed in Ref.~\cite{brs2}. 
These solutions are BPS states. Their energies have the value $4/3$, and they change to the solutions that describe the straight line orbit in the limit $r\to1/2,$ as it should be expected. 

%%%%%%%%%%%%%%%%%%%%%%%%%%%%%%%%%%%
\begin{figure}[ht]
\begin{center}
\includegraphics[height=6cm,width=10cm]{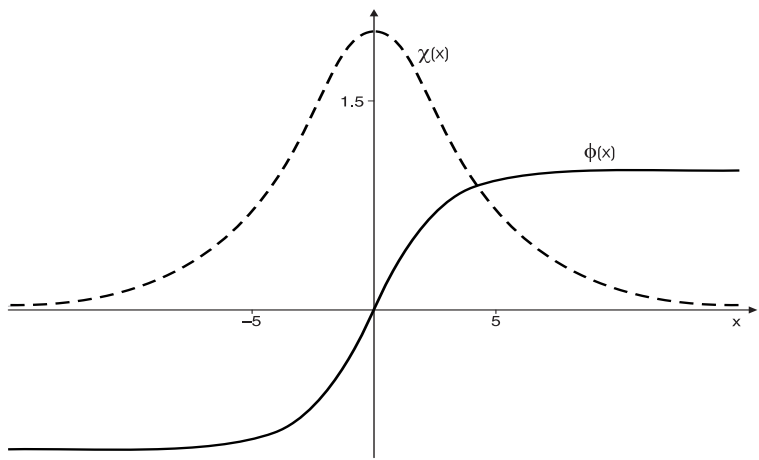}
\end{center}
\caption{A pair of solutions for the two-field model. The solid and dashed lines describe the $\phi$ and $\chi$ fields, respectively.}
\end{figure}
%%%%%%%%%%%%%%%%%%%%%%%%%%%%%%%%%%%

There are other orbits, in the other sectors, and they can be obtained as we did above. Orbits in the other sectors
were first obtained in Ref.~\cite{svo}. Also, an interesting investigation done in Ref.~\cite{spa1} obtained the integrating factor
for the pair of first-order equations, solving completely the problem of finding BPS solutions for the model.

We can study stability of the solutions explicitly. We illustrate this point investigating the simplest
case, concerning stability of the solutions for the straight line orbit in sector 1. In this case, the fluctuations in the fields $\phi$ and $\chi$ obey the equations $H\eta_n(x)=w_n^2\eta_n(x)$ and ${\wt H}\xi_n(x)={\wt w}_n^2\xi_n(x),$ respectively. The Hamiltonians are given by
\be
H=-\frac{d^2}{dx^2}+4-6\,{\rm sech}^2(x)
\ee
and
\be\label{potnbps1}
{\wt H}=-\frac{d^2}{dx^2}+4r^2-2r(1+2r)\,{\rm sech}^2(x)
\ee
The first problem is the modified P\"oschl-Teller problem. It supports two bound states, the zero-mode $w_0=0$ and $w_1=\sqrt{3}.$
The second problem is also of the modified P\"oschl-Teller type, but now the number of bound states depends on $r,$ the parameter that controls the form of the potential. We have
\be
{\wt w}_n=(4r-n)n
\ee
where $n=0,1,2,...,<2r.$ Thus, if $r\in(0,1/2],$ only the zero-mode appears. For $r\in(1/2,1],$ there are two bound states,
the zero-mode and another one, with eigenvalue ${\wt w}_1=4r-1.$ We illustrate this in Fig.~14, in which we plot the potential
of Eq.~(\ref{potnbps1}) for the values $r=1/2,$ $r=1$ e $r=3/2.$ We notice that for $\chi=0,$ the investigations for solutions reduce to that of the $\phi^4$ model; however, stability is very different, and this will certainly show up at the quantum level.

%%%%%%%%%%%%%%%%%%%%%%%%
\begin{figure}[ht]
\begin{center}
\includegraphics[height=6cm,width=10cm]{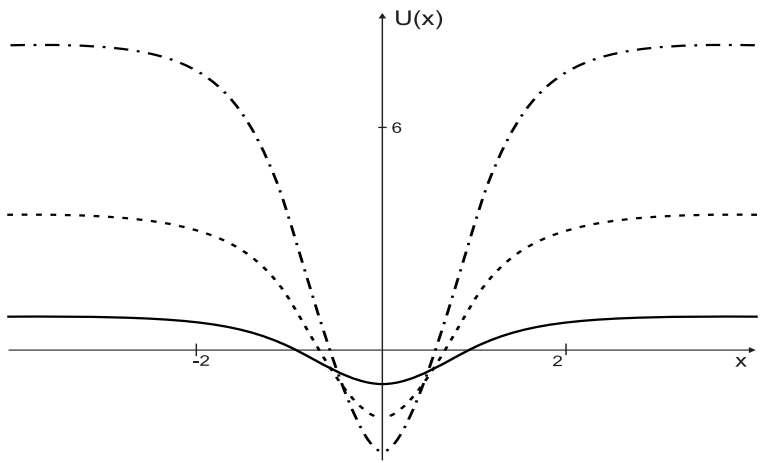}
\end{center}
\caption{Potentials for $r=1/2,$ $1,$ e $3/2,$ depicted with solid, dashed and dot-dashed lines, respectively.}
\end{figure}
%%%%%%%%%%%%%%%%%%%%%%%%%%%%%%%%

%%%%%%%%%%%%%%%%%%%%%%%%%%%%%%%%%%%%%%%%%%%%%%%%%%%%%%%%%%%
%%%%%%%%%%%%%%%%%%%%%%%%%%%%%%%%%%%%%%%%%%%%%%%%%%%%%%%%%%%
%%%%%%%%%%%%%%%%%%%%%%%%%%%%%%%%%%%%%%%%%%%%%%%%%%%%%%%%%%%
\subsection{Junction of defects}

In two-field models, in general the potential depends on the two fields, and so the minima are in the $(\phi,\chi)$ plane.
Thus, we can find situations where the minima constitute a non-collinear set of points, which is not possible for models described by a single field. For this reason, two-field models may give rise to junctions of defects, which are regions where two or more defects intersect, and this lead us to more general situations that describe junctions of defects, and the tiling of the plane with a network
of defects.  Some investigations on this subject can be found in Refs.~\cite{bb1,bb2,gt,ht,saf,oin,nol} and also in the more recent
Refs.~\cite{bms1,bms2}.

An interesting way to describe junctions of defects is with a complex field. We consider models described by the Lagrange density
\be
{\mathcal L}=\partial_\mu{\bar\varphi}\partial^\mu\varphi-V(|\varphi|)
\ee
with potential 
\be
V(|\varphi|)=|W^\prime(\varphi)|^2
\ee
given in terms of an holomorphic function $W=W(\varphi),$ with $W^\prime(\varphi)=dW/d\varphi.$
For this model, the equation of motion for static solutions has the form
\be
\frac{d^2\varphi}{dx^2}=W^\prime(\varphi)W^{\prime\prime}(\bar\varphi)
\ee
The minima of the potential are given by $W^\prime(\varphi)=0.$ We suppose that there is a finite number of minima, given by
$v_k, k=1,2,...,N,$ where $N$ is an integer. We are interested in solutions which obey the boundary conditions
\be
\varphi(x\to-\infty)\to v_i,\,\,\,\,\,\, \frac{d\varphi}{dx}(x\to-\infty)\to0
\ee 
where $v_i$ is one in the set of $N$ minima.

We can show that solutions of the first-order equation 
\be
\frac{d\varphi}{dx}=W^\prime(\bar\varphi)\,e^{i\xi}
\ee
also obey the equation of motion, for $\xi$ being real and constant.

The above model is special, since the equation of motion is equivalent to the set of first-order equations parametrized by the phase $\xi.$ To show this we use the first-order equation to write   
\be
\frac{d\varphi}{dx}=W^\prime(\bar\varphi)\,e^{i\xi}
\Rightarrow\frac{d^2\varphi}{dx^2}=W^{\prime\prime}(\bar\varphi)
\frac{d\bar\varphi}{dx}\,e^{i\xi}=
W^{\prime\prime}(\bar\varphi)W^\prime(\varphi)
\ee
This shows that solutions of the first-order equations solve the equation of motion. On the other hand, we can introduce the ratio
\be
R(\varphi)=\frac1{W^\prime(\bar\varphi)}\,\frac{d\varphi}{dx}
\ee
We use the equation of motion to get
\ben
\frac{dR}{dx}&=&\left(\frac1{W^\prime(\bar\varphi)}\right)^2
\left(W^\prime(\bar\varphi)\frac{d^2\varphi}{dx^2}-
\frac{d\varphi}{dx}W^{\prime\prime}(\bar\varphi)\frac{d{\bar\varphi}}{dx}\right)\nonumber
\\
&=&\left(\frac1{W^\prime(\bar\varphi)}\right)^2
\left(\bigl|W^\prime(\varphi)\bigr|^2-\biggl|
\frac{d\varphi}{dx}\biggr|^2\right)W^{\prime\prime}(\bar\varphi)
\een
We can also write 
\be
\frac{d}{dx}{\bigl|W^\prime(\varphi)\bigr|^2}=
W^\prime(\bar\varphi)W^{\prime\prime}(\varphi)\frac{d\varphi}{dx}+
W^\prime(\varphi)\bigl|W^{\prime\prime}(\bar\varphi)\frac{d{\bar\varphi}}{dx}
\ee
and
\be
\frac{d}{dx}\biggl|\frac{d\varphi}{dx}\biggr|^2=
W^\prime(\bar\varphi)W^{\prime\prime}(\varphi)\frac{d\varphi}{dx}+
W^\prime(\varphi)\bigl|W^{\prime\prime}(\bar\varphi)\frac{d{\bar\varphi}}{dx}
\ee
The last two equalities lead us to 
\be
S(\varphi)=\bigl|W^\prime(\varphi)\bigr|^2-\biggl|\frac{d\varphi}{dx}\biggr|^2
\ee
which does not depend on $x,$ that is, which obeys $dS(\varphi)/dx=0,$ if $\varphi$ solves the equation of motion. However, in the limit
$x\to-\infty$ the field should become $\varphi\to v_i,$ and $d\varphi/dx\to0$ and $W^\prime(\varphi)\to0.$ Thus, $S(\varphi)$ is shown to vanish, and this implies that $R(\varphi)$ is  constant. Besides, since $S(\varphi)$ vanishes, we can obtain
$\bigl|d\varphi/dx\bigr|=\bigl|W^\prime(\bar\varphi)\bigr|.$ Thus, we can write $|R(\varphi)|=1,$ and so
$R(\varphi)=e^{i\xi},$ with $\xi$ being real and constant. The last result is nothing more than an alternative way to write
the first-order equation. This is general, and it shows that in this model all the static solutions are BPS solutions.
A direct consequence of this result is that for models described by two real scalar fields, if $W=W(\phi,\chi)$ is harmonic,
then the model cannot support non-topological solutions.

We can consider the model \cite{bms1,bms2}
\be\label{wnN}
W_N^n(\varphi)=\frac1{1+n}\,\varphi^{1+n}-\frac1{1+n+N}\,\varphi^{1+n+N}
\ee
where $n=0,1,2,...,$ and $N=2,3,...$ are integers which characterize the models.
The potential has the form
\be\label{wN}
V=\frac12\,({\bar\varphi}\varphi)^n(1-\bar\varphi^N)(1-\varphi^N)
\ee
The models engender discrete symmetry, $Z_N,$ for every $n,$ and they can be used to generate defect junctions,
of intrinsic interest, and for tiling the plane with regular polygons.

The case $n=0$ is special. It has
\be
W_N(\varphi)=\varphi-\frac1{1+N}\,\varphi^{1+N}
\ee
which were already studied in \cite{fmv}, with other motivations.

The general model has minima at $v^N_k=\exp(i\xi^k_N),$ for $\xi^k_N=2\pi(k/N),$ $k=1,2,...,N,$ independently of $n,$ and also
in $v_0=0$ for $n\neq0.$ There are $N$ minima in the unit circle, and the origin is also a minimum for $n\neq0,$ with multiplicity $n.$
For the topological sectors, for $n\neq0$ there appear $N$ radial sectors. The energies of the radial sectors are given by
\be
E^n_{N,0}=\frac{N}{(1+n)(1+n+N)}
\ee
for $n=1,2,...$ and $N=2,3,...$ The energies of the other sectors are 
\be
E^n_{N,k}=2\,E^n_{N,0}\bigl|\sin[(1+n)(k\pi/N)]\bigr|
\ee
for $n=0,1,...$ and $N=2,3,...$

We notice that if the pair $n$ and $N$ obeys $n=N-1,$ the energies $E^n_{N,k}$ vanish for every $k=1,2,...,N.$ In this case the model can only support radial sectors, with energies $E^{N-1}_N=1/2N$. 

As one knows, to tile the plane with regular polygons \cite{bal,bolha}, we only have three possibilities: using equilateral
triangles, squares, or regular hexagons. Now, if we ask for the most efficient way of tiling the plane, we see that it is
given by the hexagonal tiling. Efficiency here implies the greater relation area/perimeter for the three regular polygonal tilings.
Since the use of hexagons requires the presence of triple junctions, we then focus attention on the above model, for the case $n=0$ and $N=3.$ This case can also be of interest to nanotubes and fullerenes, as explored in \cite{bb00rc,bb01}; see also
Refs.~\cite{kro,iij,dde,sdd}.

We use $n=0$ and $N=3$ to get
\be
W(\varphi)=\varphi-\frac14\varphi^4
\ee 
The three minima are $v_1=1,$ and $v_{2,3}=1/2\pm i\sqrt{3}/2.$ The energy of the defects are given by $E^0_3=3\sqrt{3}/4,$ and
this should be the energy density of the junction in the regions far away from the junction. However, if we neglect the energy density related to the junction itself, we can write the total energy of the triple junction as $E=3LE^0_3=9\sqrt{3}L/4,$ approximately,
for $L$ being the size of each junction leg. 

%%%%%%%%%%%%%%%%%%%%%%%%%%%%%%%%%%%%%%%%%%%%%%%%%%%%%%%%%%%
%%%%%%%%%%%%%%%%%%%%%%%%%%%%%%%%%%%%%%%%%%%%%%%%%%%%%%%%%%%
%%%%%%%%%%%%%%%%%%%%%%%%%%%%%%%%%%%%%%%%%%%%%%%%%%%%%%%%%%%
\section{Deformed defects}
\bigskip

In the case of models described by a single field, recently we have been able \cite{blm1,ablm} to introduce a procedure, which we call deformation procedure, where we start with a given model, that support defect structures, and we deform it, to write a new model, which also support defects that can be written in terms of the defects of the given, starting model. To make this idea practical, let us start with a model defined by the potential $V(\phi),$ which support static solutions, given by $\phi(x).$ To deform this model we get a function $f=f(\phi),$ which is well-behaved, invertible, which we name deformation function, and we write another potential, as follows
\be\label{dpot}
{\wt V}(\varphi)=\frac{V(\phi\to f(\varphi))}{[df(\varphi)/d{\varphi}]^2}
\ee 
We have two distinct models, given by
\ben
{\mathcal L}&=&\frac12\partial_\mu\phi\partial^\mu\phi-V(\phi)
\\
{\wt{\mathcal L}}&=&\frac12\partial_\mu{\varphi}\partial^\mu{\varphi}-{\wt V}({\varphi})
\een
However, if $V$ and $\wt V$ are related by Eq.~(\ref{dpot}), and if $\phi_{\pm}(x)$ are defects of the first model, then the new model
has defect solutions given by
\be\label{dd}
\varphi_\pm(x)=f^{-1}(\phi_\pm(x))
\ee 
written in terms of the inverse of the deformation function. 

To prove the above result, let us consider the equation of motion for static solutions in the two models
\ben
\frac{d^2\phi}{dx^2}=\frac{dV}{d\phi}
\\
\frac{d^2\varphi}{dx^2}=\frac{d{\wt V}}{d\varphi}
\een 
We require that the energies are finite, and that the solutions obey boundary conditions as before, in Sec.~II. Thus, we can write
\ben
\left(\frac{d\phi}{dx}\right)^2=2\,V
\\
\left(\frac{d\varphi}{dx}\right)^2=2\,{\wt V}
\een 
In the first case, the energy density has the form
\be
\varepsilon(x)=\frac12\left(\frac{d\phi}{dx}\right)^2+V\nonumber
=\left(\frac{d\phi}{dx}\right)^2
\ee 
In the other case, we get
\be
{\wt \varepsilon}(x)=\frac12\left(\frac{d\varphi}{dx}\right)^2+{\wt V}\nonumber
=\left(\frac{d\varphi}{dx}\right)^2
\ee
We can use the last expression to relate both densities
\be
{\wt \varepsilon}=\frac1{(df/d\phi)^2}\,\varepsilon(x)
\ee

On the other hand, we can use $\phi=f(\varphi)$ to obtain
\ben
\frac{d^2\phi}{dx^2}&=&\frac{d}{dx}\left(\frac{df}{d\varphi}
\frac{d\varphi}{dx}\right)\nonumber
\\
&=&\frac{df}{d\varphi}\frac{d^2\varphi}{dx^2}+
\frac{d^2f}{d\varphi^2}\left(\frac{d\varphi}{dx}\right)^2
\een
Thus, we use $d^2\varphi/dx^2$ and $d\varphi/dx$ from the equation of motion to get
\be
\frac{d^2\phi}{dx^2}=\frac{df}{d\varphi}\frac{d{\wt V}}{d\varphi}+
2\,{\wt V}\,\frac{d^2f}{d\varphi^2}
\ee
However, since ${\wt V}(\varphi)=V(\phi)/(df/d\varphi)^2,$ we can write
\be
\frac{d{\wt V}}{d\varphi}=-2\,V(\phi)\frac1{(df/d\varphi)^3}\frac{d^2f}{d\varphi^2}+
\frac1{(df/d\varphi)}\frac{dV}{d\phi}
\ee
and so we get
\be
\frac{d^2\phi}{dx^2}=\frac{dV}{d\phi}
\ee
which is the equation of motion for the first model. This result shows that solutions of the second model are related to solutions of the first model. To complete the proof, we need to show the contrary, but this is similar and straightforward.  

It is interesting to realize that we can deform the deformed model once again, etc, and this procedure lead us with an infinity of models, as we illustrate in the diagram below. 

%%%%%%%%%%%%%%%%%%%%%%%%%%%%%%%%%%%%%%%%%%%%%%%%%%%%
$$
\xymatrix{
{\cdots\widehat{\widehat{V}}} \ar[d]& &{\widehat{V}}
\ar[d] \ar[ll]_{f^{-1}} & & V
\ar[d] \ar[ll]_{f^{-1}} \ar[rr]^{f}& & {\wt{V}} \ar[d] \ar[rr]^{f} & &
{\wt{\wt{V}}\cdots} \ar[d] \\
{\cdots\widehat{\!\widehat{\phi}}}_d & &{\widehat{\phi}}_d
\ar[ll]^{f}& &\phi_d \ar[ll]^{f} \ar[rr]_{f^{-1}}& & {\wt{\phi}}_d
\ar[rr]_{f^{-1}}& & {\wt{\!\wt{\phi}}}_d \cdots}
$$
%%%%%%%%%%%%%%%%%%%%%%%%%%%%%%%%%%%%%%%%%%%%%%%%%%%%%%%%%

We notice that the deformation procedure works for models that support topological, kink-like, or non-topological, lump-like, solutions. However, for models which support BPS states, the deformation procedure is simpler, since now we can investigate the first-order equations. We have 
\be
\frac{d\phi}{dx}=\pm \frac{dW}{d\phi}
\ee
and
\be
\frac{d\varphi}{dx}=\pm\frac{1}{df/d\varphi}\frac{dW}{df}
\ee
The last equations can be written as
\be
\frac{df}{dx}=\pm\frac{dW}{df}
\ee
and now they are like the equations for the first model, with the change $\phi\to f$. Thus, their solutions obey  $f(\varphi_\pm)=\phi_\pm,$ and so we get $\varphi_\pm(x)=f^{-1}[\phi_\pm(x)],$ in accordance with Eq.~(\ref{dd}).

We illustrate the procedure with an example. We take the standard $\phi^4$ model, and the function $f(\phi)=\sinh(\phi).$ In this case the potential
\be
V(\phi)=\frac12(1-\phi^2)^2
\ee
changes to
\be
{\wt V}(\phi)=\frac12{\rm sech}^2(\phi)[1-\sinh^2(\phi)]^2
\ee
It is not hard to verify explicitly that the static solutions of this new model are given by 
\be
{\wt\phi}_\pm(x)=\pm {\rm arcsinh}[\tanh(x)]
\ee
Besides, the energy density of these solutions are 
\be
{\wt \varepsilon}(x)=\cosh^2[\tanh(x)]{\rm sech}^4(x)
\ee
where $\varepsilon(x)={\rm sech}^4(x)$ is the energy density for the $\phi^4$ model. In Figs.~15 and 16 we plot defects and energy densities in both models, for comparison.

%%%%%%%%%%%%%%%%%%%%%%%%
\begin{figure}[ht]
\begin{center}
\includegraphics[height=6cm,width=10cm]{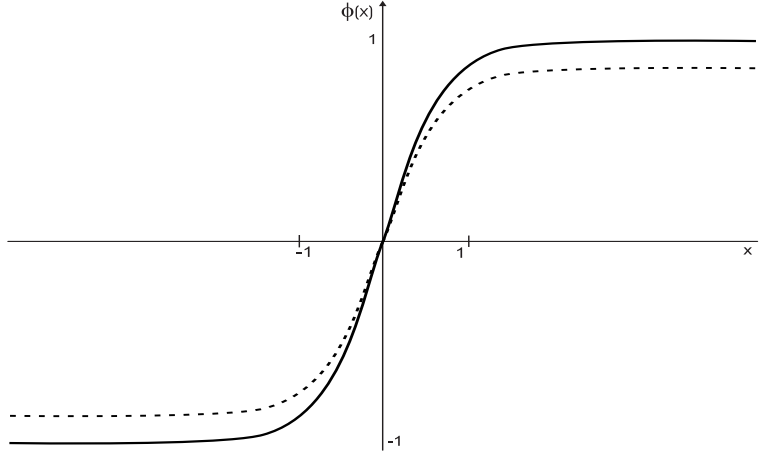}
\end{center}
\caption{Defect solutions for the $\phi^4$ model, and for the deformed model, depicted with solid and dashed lines, respectively.}
\end{figure}
%%%%%%%%%%%%%%%%%%%%%%%%%%%%%%%%

%%%%%%%%%%%%%%%%%%%%%%%%
\begin{figure}
\begin{center}
\includegraphics[height=6cm,width=10cm]{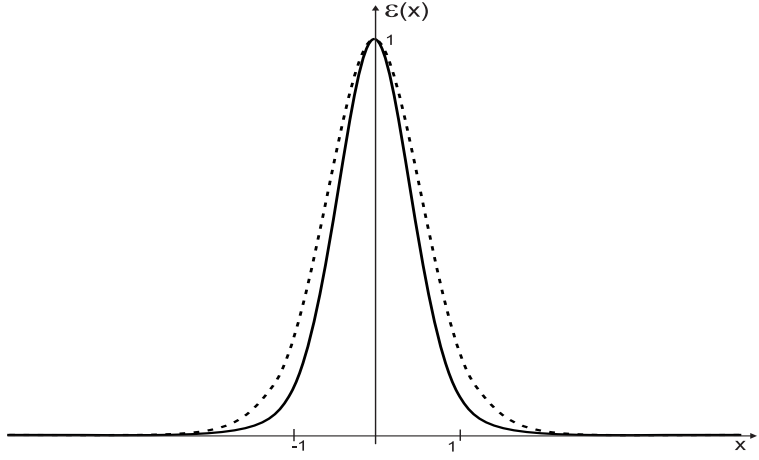}
\end{center}
\caption{Energy densities for the $\phi^4$ model, and for the deformed model, depicted with solid and dashed lines, respectively.}
\end{figure}
%%%%%%%%%%%%%%%%%%%%%%%%%%%%%%%%

Another example can be given; we consider the $\phi^4$ model, but now we use the function
\be
f_p(\varphi)=\varphi^{1/p}
\ee
where $p$ is a real parameter. We consider $p$ odd integer, $p=1,3,5,...$ The case $p=1$ is trivial, but in general the new potential has the form
\be\label{defpot}
{\wt V}(\varphi)=\frac{1}{2}p^2(\varphi^{1-1/p}-\varphi^{1+1/p})^2
\ee
This potential was already investigated in Sec.~{II}. The first-order equations are
\be
\frac{d\varphi}{dx}=\pm p(\varphi^{1-1/p}-\varphi^{1+1/p})
\ee
and their solutions can be written as 
\be
\varphi_\pm(x)=\pm \tanh^{p}(x)
\ee
as it can be shown directly.

Another example is given by the potential 
\be\label{ppi}
V(\phi)=\frac12(1-\phi^2)^3
\ee
We can use the deformation $\phi\to\tanh(\chi)$ to obtain
\be
V(\chi)=\frac12{\rm sech}^2(\chi)
\ee
Both models were investigated in Sec.~{II}. In the second case, the solutions $\chi_\pm(x)=\pm{\rm arcsinh}(x)$ can be deformed
to give 
\be
\phi_\pm(x)=\pm\frac{x}{\sqrt{1+x^2}}
\ee
which solve the first model. 

It is interesting to notice that the potential (\ref{ppi}) has no minima, and their topological solutions connect inflection points.
A similar result was presented in Ref.~\cite{jv}, where the authors find topological solutions which connect an inflection point
to a minimum of the potential there considered.

%%%%%%%%%%%%%%%%%%%%%%%%%%%%%%%%%%%%%%%%%%%%%%%%%%%%%%%%%%%
%%%%%%%%%%%%%%%%%%%%%%%%%%%%%%%%%%%%%%%%%%%%%%%%%%%%%%%%%%%
%%%%%%%%%%%%%%%%%%%%%%%%%%%%%%%%%%%%%%%%%%%%%%%%%%%%%%%%%%%
\section{Defects in higher spatial dimensions}
\bigskip

There is an old result \cite{hob,der}, which informs that models described by a single real scalar field cannot support stable solutions in space dimensions higher than $1.$ This result is known as the Derrick theorem, but we call it the Derrick-Hobart theorem, since it was also obtained by Hobart in Ref.~\cite{hob}. We can show this result as follows: in $(D,1)$ space-time dimensions, let us suppose that there is a model described by some non-negative potential, in the form
\be
{\mathcal L}=\frac12\partial_\mu\phi\partial^\mu\phi-V(\phi)
\ee 
We also suppose that $\phi(\vec{x})$ is static solution, with energy
\be
E=E_g+E_p=\int d{\vec{x}}\left(\frac12 \nabla\phi\cdot\nabla\phi+V(\phi)\right)
\ee
Now, for $\phi^\lambda(\vec{x})=\phi(\lambda\vec{x})$ we have 
\ben
E^\lambda&=&\int d{\vec{x}}\left(\frac12 \nabla\phi^\lambda\cdot\nabla\phi^\lambda+
V(\phi^\lambda)\right)\nonumber
\\
&=&
E^\lambda_g+E^\lambda_p=\lambda^{2-D}E_g+\lambda^{-D}E_p
\een
Thus, we can write
\ben
\frac{dE^{\lambda}}{d\lambda}\biggl|_{\lambda=1}&=&(2-D)\,E_g-D\,E_p
\\
\frac{d^2E^\lambda}{d\lambda^2}\biggl|_{\lambda=1}&=&(2-D)(1-D)E_g+D(1+D)E_p
\een
and so, for $E^\lambda$ to have a minimum at $\lambda=1,$ we have to have $(2-D)E_g=DE_p$ and $(2-D)(1-D)E_g+D(1+D)E_p>0,$ and this requires that $D=1.$

In spite of this result, in a recent work we have been able to circumvent the standard situation, to show how to construct
stable defect solutions for $D$ arbitrary \cite{bmm}. To make this possible, we work with the model
\be
{\mathcal L}=\frac12\partial_\mu\phi\partial^\mu\phi-U(x,\phi)
\ee
where
\be
U(x,\phi)=F(x_\mu x^\mu)V(\phi)
\ee
and $V(\phi)$ is some standard potential, and $F(x_\mu x^\mu)$ is a new function, which depends only on the space-time coordinates
in $(D,1)$ dimensions. See Ref.~\cite{G} for other investigations concerning higher dimensional defect structures.

In the case of static solutions we can choose an appropriate $F,$ to write the equation of motion in the form
\be
\nabla^2\phi=\frac1{r^{2D-2}}\frac{dV}{d\phi}
\ee
Thus, for $V=(1/2)W^2_\phi$ and for radial solutions, $\phi=\phi(r)$ should obey
\be
r^{D-1}\frac{d}{dr}\left(r^{D-1}\frac{d\phi}{dr}\right)=W_\phi W_{\phi\phi}
\ee
We can write 
\be
r^{D-1}\frac{d}{dr}=\pm\frac{d}{dy}
\ee
to change the equation of motion to 
\be
\frac{d^2\phi}{dy^2}=W_\phi W_{\phi\phi}
\ee
This equation has the form of the equation of motion for static solutions in the case of two-dimensional models. Thus, it can be written as
first-order equations 
\be
\frac{d\phi}{dy}=\pm W_\phi
\ee
or better
\be\label{bpsD}
\frac{d\phi}{dr}=\pm \frac1{r^{D-1}} W_\phi
\ee

This result can also be obtained from the energy of static, radially symmetric solutions. If we work with the energy, we can write
\ben
E&=&\frac12\Omega_D \int^{\infty}_0 r^{D-1}dr\biggl[\left(\frac{d\phi}{dr}\right)^2+
\frac1{r^{2D-2}}W^2_\phi\biggr]\nonumber
\\
&=&\pm\Omega_D\int^{\infty}_0 dr \frac{dW}{dr}+\nonumber
\\
&&\frac12\Omega_D\int_0^{\infty}r^{D-1}dr
\left(\frac{d\phi}{dr}\mp\frac1{r^{D-1}}W_\phi\right)^2
\een
where $\Omega_D=2\pi^{D/2}/\Gamma(D/2)$ stands for the angular factor, which depends on $D.$ We see that the energy is minimized to the value $E_{BPS}=\Omega_D\Delta W,$ where
\be
\Delta W=W[\phi(r\to\infty)]-W[\phi(r\to0)]
\ee
if the static, radial solution $\phi(r)$ obeys one of the two first-order equations (\ref{bpsD}).

As we have shown in Ref.~\cite{bmm}, other results can also be obtained. For instance, we can show that the radial solutions are stable
against radial fluctuations. Also, we can introduce fermions, with the standard Yukawa coupling changed to 
$Y(\phi)=r^{1-D}W_{\phi\phi},$ to include the radial modification we have introduced in the scalar potential.

On the other hand, in the case $D=2,$ the transformation which maps the problem into a one-dimensional problem has the form 
$y=\pm\ln(r)$, which very much reminds us of the Cole-Hopf transformation, used to transform the Burgers equation into a linear, exactly solved equation \cite{cole,hopf} -- see also Refs.~\cite{whi,deb}.

To illustrate the general situation, let us consider the model
\be
U(x,\phi)=\frac12\frac1{r^2}\sin^2(\phi)
\ee
in $D=2.$ This is the sine-Gordon model, but here it is modified to incorporate the spatial dependence introduced above. 
%%%%%%%%%%%%%%%%%%%%%%%%%%%%%%%%%%%%%%%%%%%%%
\begin{figure}[ht]
\begin{center}
\includegraphics[height=10cm,width=10cm]{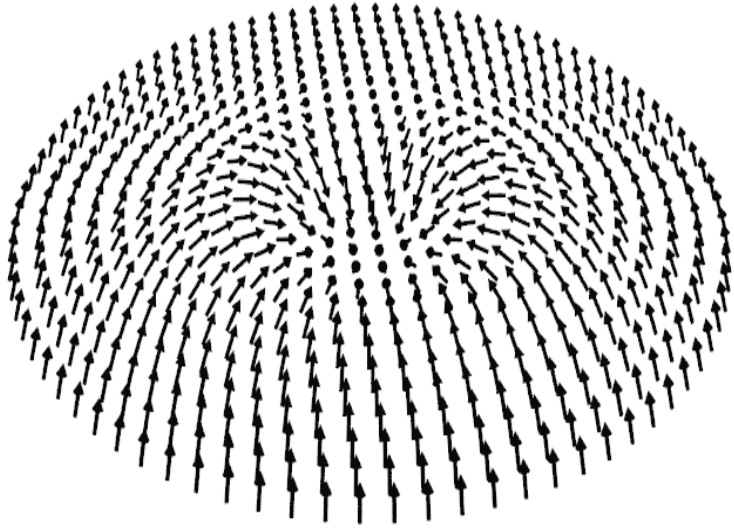}
\end{center}
\caption{Topological soliton in $D=2,$ which appears in an isotropic uniaxial ferromagnetic system.}
\end{figure}
%%%%%%%%%%%%%%%%%%%%%%%%%%%%%%%%%%%%%%%%%%%%

In this case, the equation of motion which describes static and radial solutions gets to the modified form 
\be
\frac{d^2\phi}{dr^2}+\frac1r\frac{d\phi}{dr}=\frac1{r^2}\sin(\phi)\cos(\phi)
\ee
It admits scale invariance, and it can be written as 
\be
r\frac{d}{dr}\left(r\frac{d\phi}{dr}\right)=\sin(\phi)\cos(\phi)
\ee
Thus, we take $y=\ln(r)$ to obtain
\be
\frac{d^2\phi}{dy^2}=\sin(\phi)\cos(\phi)
\ee
This problem was already solved in Sec.~{II}. We can write 
\be
\frac{d\phi}{dy}=\pm\sin(\phi)
\ee
and the solutions are
\be
\phi_{\pm}^{\pm}(y)=\pm2\arctan(e^{\pm y})
\ee
We use the radial coordinate to get
\be
\phi_{\pm}^{\pm}(r)=\pm2\arctan(r^{\pm1})
\ee
These solutions are topological solutions \cite{bmm}. They can be used to describe solitons in bidimensional, isotropic, uniaxial ferromagnetic
 \cite{kik}, and they were first described in Ref.~\cite{bpo}. In Fig.~17 we plot one of the solutions, to show the topological structure explicitly; see Ref.~\cite{brw} for related investigations. 

%%%%%%%%%%%%%%%%%%%%%%%%%%%%%%%%%%%%%%%%%%%%%%%%%%%%%%%%%%%
%%%%%%%%%%%%%%%%%%%%%%%%%%%%%%%%%%%%%%%%%%%%%%%%%%%%%%%%%%%
%%%%%%%%%%%%%%%%%%%%%%%%%%%%%%%%%%%%%%%%%%%%%%%%%%%%%%%%%%%
\section{Warped Geometry With An Extra Dimension}
\bigskip

We can couple scalar fields with gravity. We get \cite{wei,wal}
\be\label{eh}
S=\int dt d{\vec x} \sqrt{|g|}
\left(\frac14 R-\frac12g_{\mu\nu}\partial^\mu\phi\partial^\nu\phi-V(\phi)\right)
\ee
Here we are working in arbitrary $(D,1)$ space-time dimensions, using dimensionless fields and variables with $4\pi G=1.$
$g_{\mu\nu}$ is the metric tensor and $R$ is the curvature scalar, given by 
\be
R=R_\mu^{\;\;\mu}
\ee
where $R_{\mu\nu}$ is the Ricci tensor, which can be written as
\be
R_{\mu\nu}=R_{\mu\lambda\nu}^{\;\;\;\;\;\;\lambda}
\ee
in terms of the Riemann tensor $R_{\mu\nu\lambda}^{\;\;\;\;\;\;\sigma},$ which can be written in terms of the Christoffel symbols,
in the form
\be
R_{\mu\nu\lambda}^{\;\;\;\;\;\;\sigma}=\partial_\nu \Gamma^{\sigma}_{\;\;\mu\lambda}-
\partial_\mu \Gamma^{\sigma}_{\;\;\nu\lambda}+\Gamma^\alpha_{\;\;\mu\lambda}
\Gamma^\sigma_{\;\;\alpha\nu}-\Gamma^\alpha_{\;\;\nu\lambda}
\Gamma^\sigma_{\;\;\alpha\mu}
\ee
with
\be
\Gamma^\mu_{\;\;\nu\lambda}=\frac12g^{\mu\sigma}(\partial_\nu g_{\lambda\sigma}+
\partial_\lambda g_{\nu\sigma}-\partial_\sigma g_{\nu\lambda})
\ee
Einstein equation is given by
\be\label{eeq}
G_{\mu\nu}=R_{\mu\nu}-\frac12 R\,g_{\mu\nu}=2\,T_{\mu\nu}
\ee
and $G_{\mu\nu}$ is the Einstein tensor. 

Our interest on scalar fields coupled to gravity is motivated by String Theory. As one knows, superstrings need $10$ space-time dimensions
 \cite{Pol,Zwi}. In Ref.~\cite{pol1}, Polchinski has shown how to construct soliton-like brane solutions which appear under the factorization of the $(9,1)$ space-time into $AdS_5\times S^5.$ In this case, the interesting structure of space-time is $AdS_5,$ and we can work with an extra dimension, following the Randall-Sundrum \cite{RS} or Karch-Randall \cite{KR} scenarios, as we have done in Refs.~\cite{Bfg,Bg,Bbg}.

We illustrate the investigations working with the $AdS_5,$ in $(4,1)$ space-time dimensions, in the Randall-Sundrum scenario. The metric for such warped geometry with an extra dimension is given by
\be
ds^2=e^{2A}\,\eta_{\mu\nu} dx^\mu dx^\nu+dy^2
\ee
where $\eta_{\mu\nu}=(-1,1,1,1),$ and $y$ is the spatial coordinate which describes the extra dimension. $e^{2A}$ is known as the warp factor, since it modifies the standard four dimensional Minkowsky metric, described by $\eta_{\mu\nu}.$ The matter contents of the model is described by real scalar fields. We consider the action
\be
S=\int d^4x dy \sqrt{|g|}
\left(\frac14 R-\frac12g_{\mu\nu}\partial^\mu\phi\partial^\nu\phi-V(\phi)\right)
\ee
where we consider a single real scalar field. For more fields, we modify the scalar Lagrangean accordingly. 

We suppose that $A=A(y)$ and that the scalar field is also static, and only depends on the extra coordinate, $\phi=\phi(y);$ in this case the equations of motion are
\ben
&&\phi^{\prime\prime}+4A^\prime\phi^\prime=\frac{dV}{d\phi}
\\
&&A^{\prime\prime}+\frac23\phi^{\prime2}=0
\een
and
\be
A^{\prime2}-\frac16\phi^{\prime2}+\frac13V(\phi)=0
\ee
which is a consistence equation. In the above equations, we are using that $A^\prime=dA/dy,$ etc.
 
In the Randall-Sundrum scenario \cite{RS}, Minkowsky space $M_4$ is embedded in anti-de Sitter space $AdS_5.$
In Refs.~\cite{fre,new,new1} the authors introduce very interesting investigations on the subject. For instance, if we consider
the potential in the form 
\be
V(\phi)=\frac18\left(\frac{dW}{d\phi}\right)^2-\frac13W^2
\ee
we can write the first-order equations 
\be
\phi^\prime=\frac12\,\frac{dW}{d\phi}
\ee
and
\be
A^\prime=-\frac13\,W
\ee
which solve the equations of motion; see also Refs.~\cite{F1,F2} for other details on this. We notice that the first-order equation
for the scalar field is very much like the first-order equation that we get for models described by a single real scalar field in $(1,1)$ dimensions in flat spacetime; see Sec.~{II}. Thus, we can use those models to describe brane solutions in scenarios involving one extra dimension.

An example was considered in \cite{Bfg}, where we have used the model described by 
\be
W_p(\phi)=\frac{p}{2p-1}\phi^{(2p-1)/p}-\frac{p}{2p+1}\phi^{(2p+1)/p}
\ee
There we have shown that the parameter $p$ gives rise to an interesting effect, leading to thick brane with internal structure, as we show in Figs.~{18,} 19 and 20 for $p=1,3,5$. We notice that the thickness of the brane increases for increasing $p,$ and that the matter energy density opens a gap in its center, suggesting the appearance of internal structure, as we show in Ref.~\cite{Bfg}. We also notice that $p$ works like the temperature, as it is shown in the model considered in Ref.~\cite{campos}. 

%%%%%%%%%%%%%%%%%%%%%%%%%%%%%%%%%%%%%%%%%%%%%
\begin{figure}[!th]
\begin{center}
\includegraphics[angle=270,width=9cm]{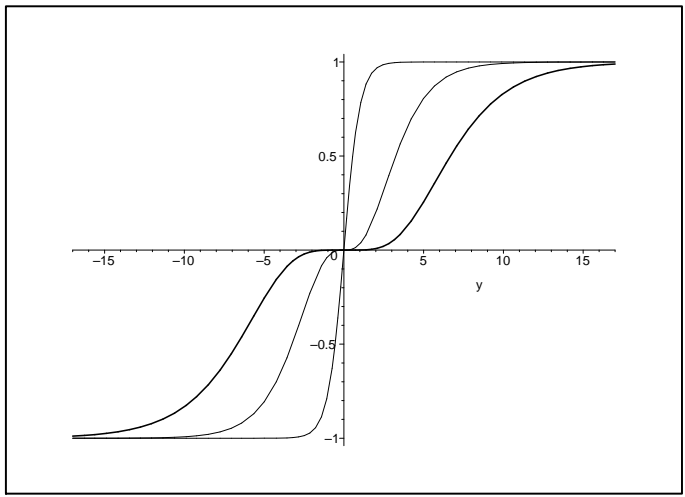}
\end{center}
\caption{The kink-like solutions for $p=1,3,5.$ The thickness of the line increases for increasing $p.$}
\end{figure}
%%%%%%%%%%%%%%%%%%%%%%%%%%%%%%%%%%%%%%%%%%%%
%%%%%%%%%%%%%%%%%%%%%%%%%%%%%%%%%%%%%%%%%%%%%
\begin{figure}[!th]
\begin{center}
\includegraphics[angle=270,width=9cm]{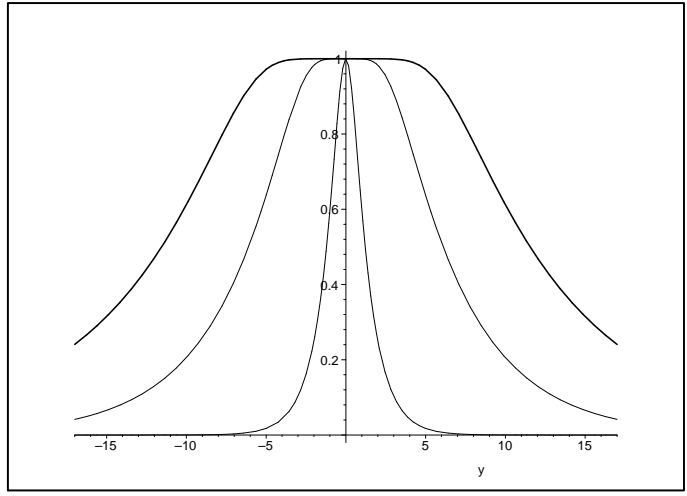}
\end{center}
\caption{The warp factor for $p=1,3,5$. The thickness of the line increases for increasing $p.$}
\end{figure}
%%%%%%%%%%%%%%%%%%%%%%%%%%%%%%%%%%%%%%%%%%%%

%%%%%%%%%%%%%%%%%%%%%%%%%%%%%%%%%%%%%%%%%%%%%
\begin{figure}
\begin{center}
\includegraphics[angle=270,width=9cm]{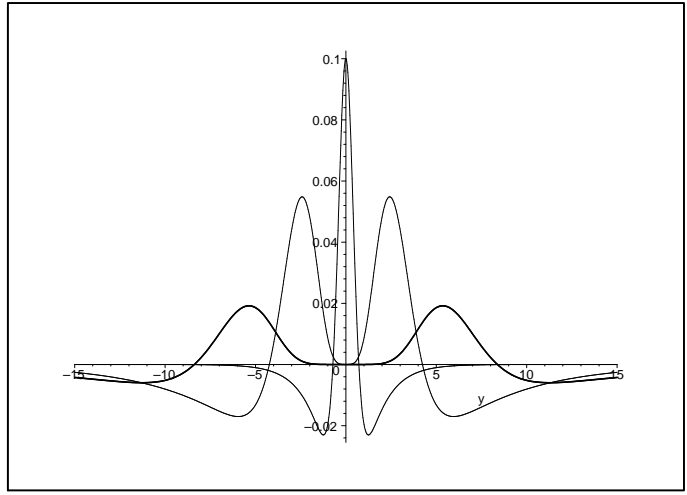}
\end{center}
\caption{The matter energy density for $p=1,3,5$. The thickness of the line increases for increasing $p.$}
\end{figure}
%%%%%%%%%%%%%%%%%%%%%%%%%%%%%%%%%%%%%%%%%%%%

In another work, we considered another model, described by \cite{Bg}
\be
W(\phi,\chi)=\phi-\frac13\phi^3-r\phi\chi^2
\ee
This model gives rise to Bloch branes, in analogy with the Bloch walls that it generates in flat space-time.
The interested reader should see Refs.~\cite{GP,melfo,AA} for other investigations.

%%%%%%%%%%%%%%%%%%%%%%%%%%%%%%%%%%%%%%%%%%%%%%%%%%%%%%%%%
%%%%%%%%%%%%%%%%%%%%%%%%%%%%%%%%%%%%%%%%%%%%%%%%%%%%%%%%%
%%%%%%%%%%%%%%%%%%%%%%%%%%%%%%%%%%%%%%%%%%%%%%%%%%%%%%%%%
\subsection{Other Scenarios}

We can work with other scenarios, involving two or more extra dimensions. In the case of two extra dimensions,
the important defect structures are vortices, which are planar topological solutions that appear in models defined in $(2,1)$
space-time dimensions, describing interactions between gauge and charged Higgs fields.
 
A model under investigation is described by \cite{bmsz}
\be
{\mathcal L}=\frac12 {\bar\kappa}\epsilon^{\mu\nu\lambda}A_\mu\partial_\nu \Box A_\lambda+
{\ov{D_\mu \varphi}}{D^\mu\varphi}-V(|\varphi|^2)
\ee
In this case, the Chern-Simons term presents higher-order derivative, and we are considering
$D_\mu=\partial_\mu-i{\bar e}\epsilon_{\mu\nu\lambda}\partial^\nu A^\lambda,$ which describes specific non-minimal coupling
between the vector and scalar fields.

In this case, the equations of motion are
\ben
\varepsilon_{\lambda\mu\nu}\partial^\mu ({\bar\kappa}\Box A^\nu+J^\nu)&=&0
\\
D_\mu D^\mu\varphi+\frac{\partial V}{\partial {\bar\varphi}}&=&0
\een
where
\be
J^\mu=i{\bar e}[{\bar\varphi}D^\mu\varphi-(\overline{D^\mu\varphi})\varphi]
\ee
is the Noether current.

The energy density for static configurations can be written in the form
\be
\varepsilon=|{\bf D}\varphi|^2+
\frac{{\bar\kappa}^2}{4{\bar e}^2}\frac{(\nabla^2A_0)^2}{|\varphi|^2}+V(|\varphi|^2)
\ee
where ${\bf D}={\vec\nabla+i{\bar e}{\vec n}\times{\vec\nabla}A_0}.$ This result appears after eliminating the constraint
\be
B=-\frac{\bar\kappa}{2{\bar e}^2}\frac{\nabla^2 A_0}{|\varphi|^2}
\ee
The energy gets the form
\ben
E&=&2\pi \int rdr\left(|(D_1\pm iD_2)\varphi|^2+\Bigl|\frac{\bar\kappa}{2{\bar e}}
\frac{\nabla^2A_0}{\varphi}\mp\frac{{\bar e}^2}{\lambda}{\bar\varphi}
(1-|\varphi|^2)\Bigr|^2\right)\nonumber
\\
&&\pm2\pi\int r dr {\nabla^2A_0}
\een
where we have set
\be
V(|\varphi|^2)=\frac{{\bar e}^4}{{\bar\kappa}^2}|\varphi|^2(1-|\varphi|^2)^2
\ee
The energy can be minimized to 
\be
E_{BPS}=\pm 2\pi \int r dr\nabla^2 A_0
\ee
It is obtained for fields that obey the first-order equations 
\ben
&&D_1\varphi=\mp iD_2\varphi
\\
&&B=\mp\frac{{\bar e}}{\bar\kappa}(1-|\varphi|^2)
\een

We choose the fields in the form 
\ben
\varphi&=&g(r) e^{in\theta}
\\
\frac{dA_0}{dr}&=&-\frac1{{\bar e}r}[a_0(r)-n]
\een
where $n=\pm1,\pm2,...,$ The boundary conditions are 
\ben
g(r\to0)&\to&0.\;\;\;g(r\to\infty)\to1
\\
a_0(r\to0)&\to& n,\;\;\;a_0(r\to\infty)\to 0
\een
Thus, the first-order equations change to 
\ben
\frac{dg}{dr}&=&\pm\frac1r g a_0
\\
\frac1r\frac{da_0}{dr}&=&\mp2\frac{{\bar e}^4}{{\bar\kappa}^2} g^2(1-g^2)
\een
The energy is quantized, in the form $E_{BPS}=2\pi|n|.$

This models contains the Chern-Simons term; thus, $J^0=-2{\bar e}^2|\varphi|^2B$ shows that the electric charge is given by
$Q=2\pi ({\bar\kappa}/{\bar e})n.$ On the other hand, the magnetic flux is given by $\Phi_B=\pm(\pi{\bar\kappa}/{\bar e}^3)(n^2+|n|).$ Moreover, if we introduce the topological current $J^\mu_T=\varepsilon^{\mu\nu\lambda}\partial_\nu A_\lambda,$ we get that
$Q_T=\pm(\pi {\bar\kappa}/{\bar e}^3)(n^2+|n|),$ which can be identified with the magnetic flux. This identification also appears
in the standard Chern-Simons model, but there the values are different \cite{hkp,jw}: $\Phi_B=Q_T=\pm(2\pi/e)n.$

Here, however, we can introduce another topological current, dual to the above one. It has the form
${\wt J}^\mu_T=\varepsilon^\mu_{\,\,\nu\lambda}\partial^\nu J^\lambda_T.$ In the standard Chern-Simons model, the corresponding topological charge vanishes. But in the model under investigation, it is given by ${\wt Q}_T=\pm2\pi({\bar\kappa}/{\bar e})n,$ which is exactly the electric charge of the solution.

Our interest in vortices is related to the interest in braneworld scenarios with two or more extra dimensions, but this is another story.

%%%%%%%%%%%%%%%%%%%%%%%%%%%%%%%%%%%%%%%%%%%%%%%%%%
%%%%%%%%%%%%%%%%%%%%%%%%%%%%%%%%%%%%%%%%%%%%%%%%%%
%%%%%%%%%%%%%%%%%%%%%%%%%%%%%%%%%%%%%%%%%%%%%%%%%%
\section*{Acknowledgments}

The author would like to thank the Organizing Committee of the XIII J.A. Swieca Summer School on Particles and Fields for the kind invitation, and F.A. Brito, C. Furtado, A.R. Gomes, L. Losano, J.M.C. Malbouisson, R. Menezes, J.R. Nascimento, R.F. Ribeiro, and C. Wotzasek for their comments and suggestions. He also thanks CAPES, CNPq, PROCAD/CAPES and PRONEX/CNPq/FAPESQ for financial support.

%%%%%%%%%%%%%%%%%%%%%%%%%%%%%%%%%%%%%%%%%%%%%%%%%%%%%%%%%%%%%
%%%%%%%%%%%%%%%%%%%%%%%%%%%%%%%%%%%%%%%%%%%%%%%%%%%%%%%%%%%%%

\end{document}